\documentclass[twocolumn,aps,prb,superscriptaddress]{revtex4-2}

\usepackage{amsmath}
\usepackage{amssymb}
\usepackage{graphicx}
\usepackage{color}

\newcommand{\be}{\begin{equation}}
\newcommand{\ee}{\end{equation}}
\newcommand{\bea}{\begin{eqnarray}}
\newcommand{\eea}{\end{eqnarray}}
\newcommand{\bse}{\begin{subequations}}
\newcommand{\ese}{\end{subequations}}
\newcommand{\sma}{${\rm SrMn_2As_2}$}
\newcommand{\bma}{${\rm BaMn_2As_2}$}
\newcommand{\cma}{${\rm CaMn_2As_2}$}
\newcommand{\bms}{${\rm BaMn_2Sb_2}$}
\newcommand{\sms}{${\rm SrMn_2Sb_2}$}
\newcommand{\smp}{${\rm SrMn_2P_2}$}

\newcommand{\cmp}{${\rm CaMn_2P_2}$}
\newcommand{\cas}{${\rm CaAl_2Si_2}$}
\newcommand{\tcs}{${\rm ThCr_2Si_2}$}

\newcommand{\scmp}{${\rm (Sr,Ca)Mn_2P_2}$}

\begin{document}

\title{First-order transitions at the N\'eel temperatures of trigonal SrMn$_2$P$_2$ and CaMn$_2$P$_2$ single crystals containing corrugated-honeycomb Mn sublattices}

\author{N. S. Sangeetha}
\altaffiliation{Present address: Institute for Experimental Physics~IV, Ruhr University Bochum, 44801 Bochum, Germany}
\affiliation{Ames Laboratory, Iowa State University, Ames, Iowa 50011, USA}
\author{Santanu Pakhira}
\affiliation{Ames Laboratory, Iowa State University, Ames, Iowa 50011, USA}
\author{Q.-P. Ding}
\affiliation{Ames Laboratory, Iowa State University, Ames, Iowa 50011, USA}
\affiliation{Department of Physics and Astronomy, Iowa State University, Ames, Iowa 50011, USA}
\author{H.-C. Lee}
\affiliation{Ames Laboratory, Iowa State University, Ames, Iowa 50011, USA}
\affiliation{Department of Physics and Astronomy, Iowa State University, Ames, Iowa 50011, USA}
\author{V.~Smetana}
\author{A.-V. Mudring}\affiliation{Department of Materials and Environmental Chemistry, Stockholm University, Svante Arrhenius v\"ag 16 C, 106 91 Stockholm, Sweden}
\author{Y.~Furukawa}
\affiliation{Ames Laboratory, Iowa State University, Ames, Iowa 50011, USA}
\affiliation{Department of Physics and Astronomy, Iowa State University, Ames, Iowa 50011, USA}
\author{D. C. Johnston}
\affiliation{Ames Laboratory, Iowa State University, Ames, Iowa 50011, USA}
\affiliation{Department of Physics and Astronomy, Iowa State University, Ames, Iowa 50011, USA}

\date{\today}

\begin{abstract}

Single crystals of  \cmp\ and \smp\  were grown using Sn flux and characterized by room-temperature single-crystal x-ray diffraction, electrical resistivity~$\rho$, heat capacity~$C_{\rm p}$, and magnetic susceptibility $\chi=M/H$ measurements versus temperature~$T$ and magnetization $M$ versus applied magnetic field~$H$ isotherm measurements. The x-ray diffraction results show that both \smp\ and \cmp\ adopt the trigonal \cas-type structure. The $\rho(T)$ measurements demonstrate insulating ground states for both compounds with intrinsic activation energies of 0.124~eV for \smp\ and 0.088~eV for \cmp\@. The $\chi(T)$ and  $C_{\rm p}(T)$ data reveal a weak first-order antiferromagnetic (AFM) transition at the  N\'eel temperature $T\rm_N$ = 53(1)~K for \smp\@ and a strong first-order AFM transition at $T_{\rm N} =69.8(3)$~K for \cmp.  Both compounds show an isotropic and nearly $T$-independent $\chi(T\leq T{\rm_N})$\@.  $^{31}$P NMR measurements confirm the strong first-order transition in \cmp\ but show critical slowing down near $T_{\rm N}$ for \smp\ thus evidencing second-order character.  The NMR measurements also indicate that the AFM structure of \cmp\ is commensurate with the lattice whereas that of \smp\ is incommensurate.  These first-order AFM transitions are unique among the class of trigonal (Ca, Sr, Ba)Mn$_2$(P, As, Sb, Bi)$_2$ compounds which otherwise exhibit second-order AFM transitions.  This result presents a challenge to understand the systematics of magnetic ordering in this class of materials in which magnetically-frustrated antiferromagnetism is quasi-two-dimensional.

\end{abstract}

\maketitle

\section{Introduction}

The Mn-based 122-type pnictides $A{\rm Mn_2}Pn_2$ ($A=$ Ca, Sr, Ba; $Pn$ = P, As, Sb, Bi) have received attention owing to their close structural relationship to high-$T\rm_c$ iron pnictides. The undoped Mn pnictides are local-moment antiferromagnetic (AFM) insulators like the high-$T_{\rm c}$ cuprate parent compounds \cite{{An2009}, {Singh2009}, {Johnston2011}}.  The ${\rm BaMn_2}Pn_2$ compounds crystallize in the body-centered tetragonal \tcs\ structure as in $A\rm Fe_2As_2$ ($A$ = Ca, Sr, Ba, Eu), whereas the ${\rm (Ca,Sr)Mn}_2Pn_2$ compounds crystallize in the trigonal \cas-type structure \cite{Brechtel1979}.   Recently, density-functional theory (DFT) calculations for the 122 pnictide family have suggested that the trigonal 122 transition-metal pnictides which have the \cas\ structure might comprise a new family of magnetically-frustrated materials in which to study the potential superconducting mechanism \cite{Zeng2017, Fouet2001}.  It had previously been suggested on theoretical grounds that CaMn$_2$Sb$_2$ is a fully-frustrated classical magnetic system arising from proximity to a tricritical point~\cite{Mazin2013, Simonson2012, McNally2015}.

The electrical resistivity $\rho$ and heat capacity $C_{\rm p}$ versus temperature~$T$ of single-crystal \cmp\ were reported in Ref.~\cite{Li2020}. The compound is an insulator at $T=0$ and undergoes a first-order transition of some type at 69.5~K\@. Raman spectra of \cmp\ versus~$T$ suggested the formation of a crystallographic superstructure below 69.5~K\@.  However, the authors' magnetic susceptibility $\chi(T)$ measurements revealed no magnetic transition at that temperature.  Previous studies on polycrystalline \smp\ showed that it is an insulator at room temperature with AFM ordering at $T{\rm_N}=53(1)$~K \cite{Brock1994}. \smp\ transforms from its trigonal \cas-type structure at atmospheric pressure~$p$ to the tetragonal \tcs\ structure (space group $I4/mmm$) after treatment at $p= 5$~GPa and $T= 900\,^\circ$C~\cite{Xie2017}. 
 
Here we report the detailed properties of trigonal Mn pnictides \cmp\ and  \smp~\cite{Mewis1978} single crystals.  We present the results of single-crystal x-ray diffraction, electrical resistivity $\rho$ in the $ab$ plane versus temperature $T$, isothermal magnetization versus applied magnetic field $M(H)$, magnetic susceptibility $\chi(T)$, heat capacity $C_{\rm p}(H,T)$, and $^{31}$P NMR measurements.  We find from $C_{\rm p}(T)$, $\chi(T)$, and NMR that \cmp\ exhibits a strong first-order AFM transition at $T_{\rm N} = 69.8(3)$~K whereas \smp\ shows a weak first-order transition at $T_{\rm N} = 53(1)$~K but with critical slowing down on approaching $T_{\rm N}$ from above as revealed from NMR, a characteristic feature of second-order transitions. Thus the AFM transition in \smp\ has characteristics of both first- and second-order transitions.  The $\chi(T)$ data also reveal the presence of strong isotropic AFM spin fluctuations in the paramagnetic (PM) state above $T_{\rm N}$ up to our maximum measurement temperatures of 900~K and 350~K for \smp\ and \cmp, respectively, likely arising from the quasi-two-dimensional nature of the Mn spin layers~\cite{Lines1969} together with possible contributions from magnetic frustration.

Remarkably, our studies of \smp\ and \cmp\ reveal the only known members of the isostructural trigonal class of materials with general formula $A$Mn$_2Pn_2$ containing Mn$^{2+}$ spins $S=5/2$ that exhibit first-order AFM transitions, where $A$ = Ca, Sr, or Ba and the pnictogen $Pn=$~P, As, Sb, or Bi.  The others show second-order AFM transitions, including ${\rm CaMn_2As_2}$~\cite{Sangeetha2016}, ${\rm SrMn_2As_2}$~\cite{Brock1994, Sangeetha2016, Das2017}, ${\rm CaMn_2Sb_2}$~\cite{Bobev2006, Ratcliff2009, Bridges2009, Simonson2012, McNally2015, Sangeetha2018}, ${\rm SrMn_2Sb_2}$~\cite{Bobev2006, Sangeetha2018}, and ${\rm CaMn_2Bi_2}$~\cite{Gibson2015}.  The AFM transition in the monoclinic compound ${\rm Li_2MnO_3}$ containing a Mn honeycomb lattice with Mn$^{4+}$ spins $S=3/2$ is also second order~\cite{Lee2012}. 

Following the experimental details in Sec.~\ref{ExpDetails}, the single-crystal structure data for single-crystal \smp\ and \cmp\ are given in Sec.~\ref{Sec:Struct}, the $\rho(T)$ data in Sec.~\ref{Sec:Rho}, our studies of $M(H)$ and $\chi(T)$ in Sec.~\ref{Sec:MandChi}, the $C_{\rm p}(T)$ measurements  in Sec.~\ref{Sec:HC}, the NMR results in Sec.~\ref{NMR}.  Concluding remarks are given in Sec.~\ref{Sec:Summary}.

\section{\label{ExpDetails} Experimental Details}

Single crystals of \smp\ and \cmp\ were grown in Sn flux. High-purity elements Sr (99.99\%) from Sigma Aldrich and Ca (99.98\%), Mn (99.95\%), P (99.999\%), and Sn (99.999\%) from Alfa Aesar were taken in the molar ratio (Sr,Ca):Mn:P:Sn = 1.05:2:2:20 and placed in an alumina crucible that was sealed under $\approx 1/4$ atm Ar pressure in a silica tube. Excess Sr or Ca was used in the synthesis to avoid the formation of MnP and thus to enhance the formation of magnetically-pure samples. After preheating the mixture at 600~$^{\circ}$C for 7~h, the assembly was heated to 1150~$^{\circ}$C at a rate of 50~$^{\circ}$C/h and held at this temperature for 15~h for homogenization. Then the furnace was cooled at a rate of 4~$^{\circ}$C/h to 700~$^{\circ}$C\@. Shiny platelike hexagon-shaped single crystals of \smp\ and \cmp\ were obtained after decanting the Sn flux using a centrifuge.  

Chemical analyses of the single crystals were performed using a JEOL scanning-electron microscope (SEM) equipped with an EDX (energy-dispersive x-ray analysis) detector, where a counting time of 120 s was used. Single-crystal x-ray structural analyses of \smp\ and \cmp\ were performed at room temperature using a Bruker D8 Venture diffractometer equipped with a Photon 100 CMOS detector, a flat graphite monochromator and a Mo~${\rm K_\alpha}$ I$\mu$S microfocus x-ray source \mbox{($\lambda = 0.71073$~\AA)} operating at a voltage of 50~kV and a current of 1~mA\@. The raw frame data were collected using the Bruker APEX3 software package  \cite{APEX2015}, while the frames were integrated with the Bruker SAINT program \cite{SAINT2015} using a narrow-frame algorithm integration of the data and were corrected for absorption effects using the multiscan method (SADABS)~\cite{Krause2015}.  The atomic thermal factors were refined anisotropically.  Initial models of the crystal structures were first obtained with the program SHELXT-2014~\cite{Sheldrick2015A} and refined using the program SHELXL-2014~\cite{Sheldrick2015C} within the APEX3 software package.

Magnetic susceptibility $\chi=M(T)/H$ at fixed applied magnetic field~$H$ over the $T$ range $1.8 \leq T \leq 350$~K and $M(H)$ isotherm measurements for $H \leq 5.5$~T were carried out using a Quantum Design, Inc., Magnetic Properties Measurement System (MPMS). The high-temperature $M(T)$ for $300 \leq T \leq 900$~K was measured  using the vibrating sample magnetometer (VSM) option of a Quantum Design, Inc., Physical Properties Measurement System (PPMS)\@.  Four-probe dc $\rho(T)$ and $C{\rm_p}(T)$ measurements were carried out on the PPMS, where electrical contacts to a crystal for the $\rho(T)$ measurements were made using annealed 0.05~mm-diameter Pt wires attached to the crystals with silver epoxy.

\begin{figure}
\includegraphics[width=3.3in]{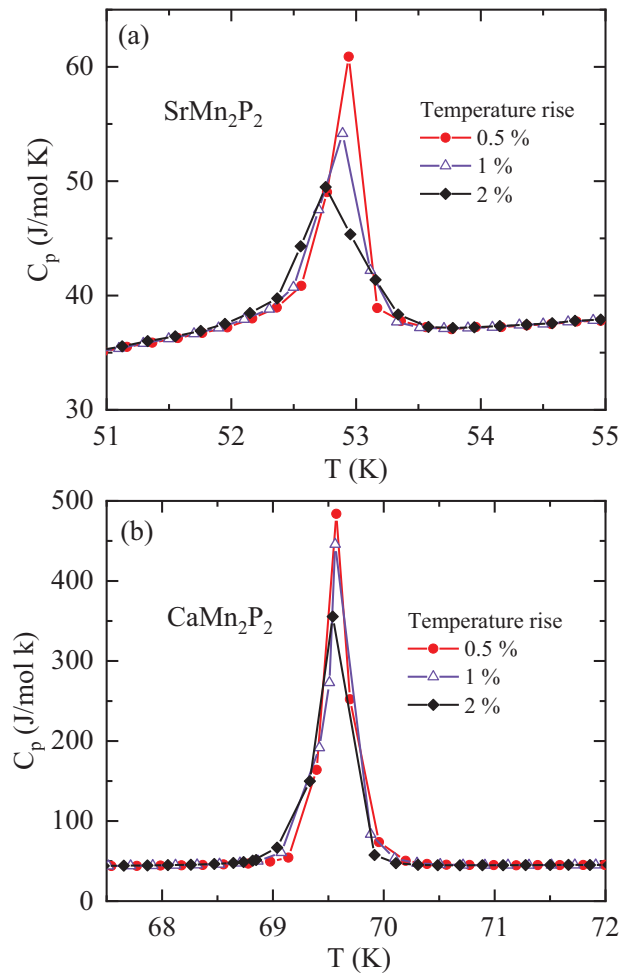}
\protect\caption{Measured $C{\rm_p}(T)$ data of (a) \smp\ and (b) \cmp\ single crystals around FOMT for different temperature-rise values using standard semi-adiabatic method implemented in the PPMS system by Quantum Design.}
\label{Heat_Cap_Trise}
\end{figure}

The heat-capacity measurements were carried out using the standard semiadiabatic method implemented in the PPMS system by Quantum Design. This technique is widely followed for a second-order magnetic phase transitions and the relative accuracy of the results are within 1\%. However, in the present study of thermal hysteresis and latent heat of a first-order magnetic transition, this method does not provide the most accurate results. Therefore, the $C{\rm_p}(T)$ values associated with the first-order transitions found in \cmp\ and \smp\ were estimated using the single-curve slope analysis method described in Refs.~\cite{Hardy2009, Suzuki2010, Guillou2018, PPMSQD}. When the sample temperature crosses a first-order transition, the temperature versus time response curve $T(t)$ allows $C_{\rm p}$ at temperature~$T$ to be found from the $T(t)$ data using the relation
\be
\ C_{\rm p}(T) = \frac{-K_{\rm w}(T-T_{\rm p}) + P(T)}{\frac{dT}{dt}(T)},
\label{slopeanalysis}
\ee
where, $K_{\rm w}$ is the thermal conductance of the supporting wires of the sample platform in the heat-capacity puck, $T$ is the temperature of the platform, $T_{\rm p}$ is the temperature of the heat-capacity puck (system temperature), and $P$ is the heater power. The recent version of MultiVu software is implemented with the single-heat-pulse slope analysis as described in Section 4.6 of Ref.~\cite{PPMSQD}. However, to obtain an accurate value for the latent heat using the slope analysis method, the thermal coupling between the sample and the platform needs to be excellent which can be obtained by applying a very small amount of grease for thermal contact and using a flat thin crystal. The temperature rise of the measurement should be smaller than the width of the peak; otherwise, the heat-capacity peak is artificially broadened. We optimize the temperature rise parameter to obtain the sharpest peak associated with the transition. Figure~\ref{Heat_Cap_Trise} depicts the measured $C{\rm_p}(T)$ data around the first order transition for different temperature rise values.

    NMR measurements of $^{31}$P  nuclei with nuclear spin $I$ = 1/2 and gyromagnetic ratio $\gamma$$_{\rm N}$/2$\pi$ = 17.237 MHz/T were conducted using a lab-built phase-coherent spin-echo pulse spectrometer. 
   Single crystals of SrMn$_2$P$_2$ ($4\times3\times0.1$ mm$^3$) and CaMn$_2$P$_2$  ($3\times3\times0.1$ mm$^3$) were used for NMR  measurements.
   $^{31}$P NMR spectra were obtained either by fast Fourier transform (FFT) of the NMR echo signals under an  external magnetic field   of \mbox{$\sim$ 7.01~T} or by sweeping the external magnetic field  $H$ at a constant resonance frequency of 121 MHz.
 The $^{31}$P spin-lattice relaxation rate $(1/T_{1})$ was measured  with a saturation recovery method.
   $1/T_1$ at each $T$ was determined by fitting the nuclear magnetization $M$ versus time $t$  using the streched exponential function $1-M(t)/M(\infty) = e^{(-t/T_{1})^\beta}$,  where $M(t)$ and $M(\infty)$ are the nuclear magnetization at time $t$ after the saturation and the equilibrium nuclear magnetization at $t$ $\rightarrow$ $\infty$, respectively. 
  All the relaxation data were well fitted with the streched exponent $\beta$ = 1 in the paramagnetic state for both samples.
  In the AFM state, $\beta \sim 0.6 \pm 0.15$ was used for SrMn$_2$P$_2$.
  For CaMn$_2$P$_2$ in the AFM state, $\beta = 1$ was used.
  However, at low temperatures below $\sim 20$~K, we observed a slight deviation from the function and fitted with two $T_1$ components and took the long component as $T_1$. 

\section{\label{Sec:Struct} Crystal Structure}

\begin{figure}
\includegraphics[width=3.3in]{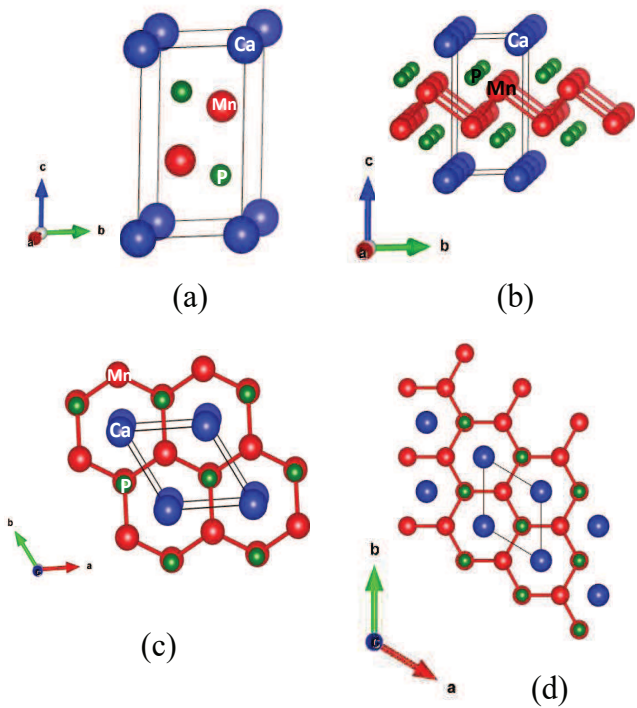}
\caption{Crystal structure of \cmp.  (a)~One unit cell of the trigonal structure. The Mn atoms are represented by filled red spheres, the phosphorus atoms by smaller filled green spheres, and the Ca atoms by larger filled blue spheres.  (b)~Projection of the structure on the $b$-$c$ plane (of the hexagonal unit cell), showing the corrugated nature of the Mn honeycomb sublattice.  (c,d)~Projections of the lattice onto the $a$-$b$ plane, emphasizing the projected corrugated honeycomb Mn sublattice.}
\label{Fig:CaMn2P2_Unitcell}
\end{figure}

\begin{table}
\caption{\label{Table_SXRD} Refined crystallographic parameters obtained from single XRD of \smp\ and \cmp\ crystals. The atomic coordinates in \scmp\ of the hexagonal unit cell are Sr/Ca: 1a (0, 0, 0); Mn: 2d (1/3, 2/3, $z \rm_{Mn}$); and P: 2d (1/3, 2/3, $z{\rm_{P}}$).}

\begin{tabular}{p{3.5cm}|p{2.5cm}|p{2.3cm}}
\hline
\hline 
\\
 & \smp\  & \cmp\ \\
\hline
Structure  & \cas-type trigonal &   \cas-type trigonal \\

Space group  & $P\bar{3}m1$  & $P\bar{3}m1$ \\
\\
Lattice parameters &  &   \\
\hspace{0.8cm} $a$ (\AA )  & 4.656(6)  &   4.1013(3)\\
\hspace{0.8cm} $c$ (\AA )  & 7.138(1)  &  6.8564(6)\\
\hspace{0.8cm} $c/a$  	  & 1.7136(5)  &  1.6717(3)\\
\hspace{0.8cm} $V_{{\rm cell}}$ (\AA $^{3}$)  & 107.27(4)  &  99.88(2) \\
Atomic coordinates & & \\
\hspace{0.8cm} $z{\rm_{Mn}}$  & 0.62033(7)  &  0.62441(4)\\
\hspace{0.8cm} $z{\rm_{P}}$   & 0.2726(1) &  0.26155(9)  \\
\hline

\end{tabular}
\end{table}

SEM images of the crystal surfaces indicated single-phase crystals. EDX analyses of the chemical compositions were in agreement with the expected 1:2:2 stoichiometry of the compounds and the amount of Sn incorporated into the crystal structure from the Sn flux is zero to within the experimental error.

Single-crystal XRD measurements on $\rm{CaMn_2P_2}$ and $\rm{SrMn_2P_2}$ confirmed the single-phase nature of the compounds and the \cas-type crystal structure of each that is shown in Fig.~\ref{Fig:CaMn2P2_Unitcell}.   The crystal data are listed in Table~\ref{Table_SXRD}. The lattice parameters $a$ and $c$ of the trigonal structure (hexagonal unit cell) are in good agreement with previous values~\cite{Mewis1978, Li2020}.  The crystal structure of \cmp\ at 293 and 40~K was found to be the same~\cite{Li2020}. However, as noted in the Introduction, Raman spectra of \cmp\ versus~$T$ suggested the formation of a crystallographic superstructure below 69.5~K, which was stated to be consistent with the authors' single-crystal x-ray diffraction data~\cite{Li2020}.

\section{\label{Sec:Rho}  Electrical Resistivity}

\begin{figure}
\includegraphics[width=3.3in]{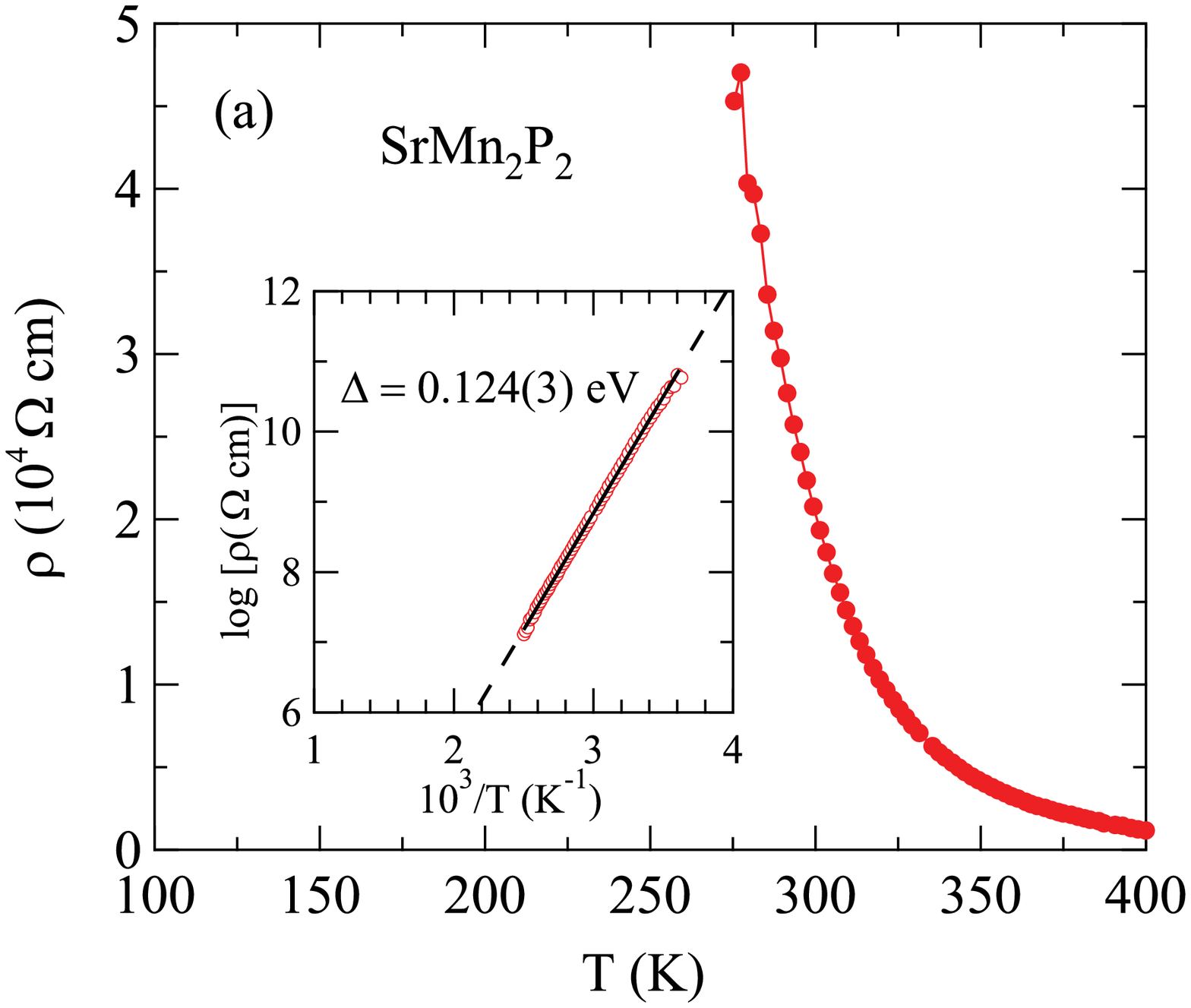}
\includegraphics[width=3.3in]{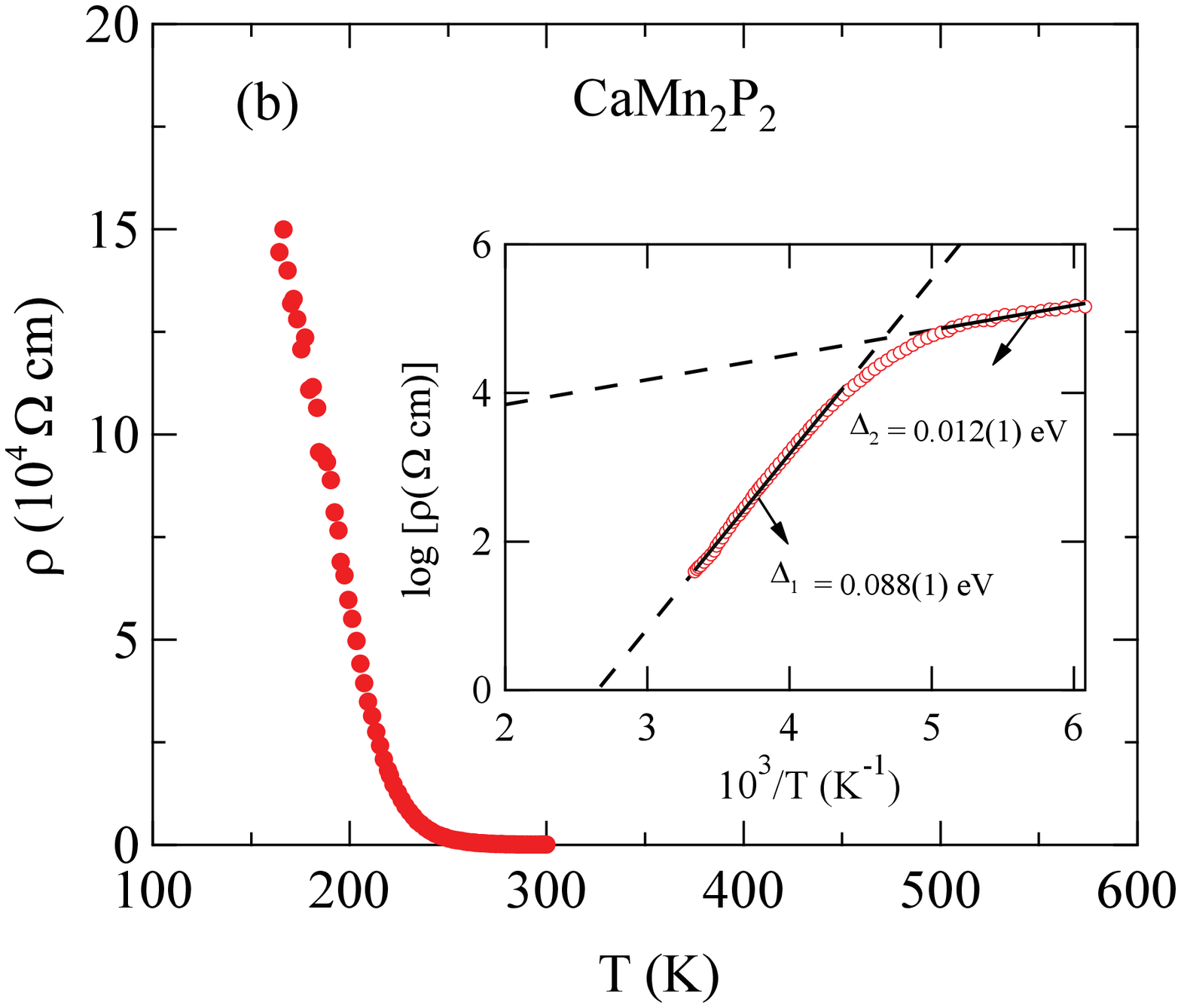}
\caption{Temperature $T$ dependence of the electrical resistivity, $\rho(T)$, in the $ab$~plane for (a)~\smp\ and (b)~\cmp. The insets show plots of $\log_{10} \rho$ versus $1/T$\@. The solid straight lines through the inset data are fits over restricted temperature intervals by Eq.~(\ref{rho_fit}) as discussed in the text, and the dashed lines are extrapolations.}
\label{Fig:scmpRho}
\end{figure}

The $ab$-plane $\rho(T)$ data from 270 to 400~K for \smp\ and from 150 to 300~K for \cmp\ are shown in  Figs.~\ref{Fig:scmpRho}(a) and \ref{Fig:scmpRho}(b), respectively.  For both compounds $\rho$ first increases slowly with decreasing~$T$  and then increases more rapidly.   The data demonstrate that both \smp\ and \cmp\ have insulating ground states. We fitted the respective $\rho(T)$ data over restricted temperature intervals by
\be
\log_{10}\rho = A + 2.303\left(\frac{\Delta}{k{\rm_B}T}\right),
\label{rho_fit}
\ee
where $A$ is a constant, $k\rm_B$ is Boltzmann's constant, and $\Delta$ is the activation energy.

Plots of $\log_{10} \rho$ vs $1/T$  are shown in the insets of Figs.~\ref{Fig:scmpRho}(a) and~\ref{Fig:scmpRho}(b) for \smp\ and \cmp,\ respectively. For \smp, the high-$T$ data between  $\approx 270$~K and 400~K are nearly linear in $1/T$ and were fitted by Eq.~(\ref{rho_fit}), yielding the intrinsic activation energy \mbox{$\Delta =  0.124(3)$~eV} as shown by the solid straight line in the inset of Fig.~\ref{Fig:scmpRho}(a), and the extrapolations as dashed lines. The fitted activation energy is of the same order as found previously for isostructural \bms\ \cite{Sangeetha2018}. However, the value here is much larger than the activation energy found for polycrystalline \smp\ over the temperature range 200--300~K, which was  0.0129(2)~eV~\cite{Brock1994}.

For \cmp, the data between 220~K and 300~K and between 197~K and 164~K were respectively nearly linear in $1/T$ and were fitted by Eq~(\ref{rho_fit}), yielding the intrinsic activation energy at high~$T$ as $\Delta = 0.088(1)$~eV, whereas the extrinsic activation energy at low~$T$ is \mbox{$\Delta = 0.012(1)$~eV}\@.  The fits are shown in the inset of Fig.~\ref{Fig:scmpRho}(b) as the solid straight lines and the extrapolations by dashed lines. For comparison, the high-$T$ intrinsic activation energy reported recently for single-crystal \cmp\ from $ab$-plane $\rho(T)$ data was $\Delta = 0.040$~eV and the extrinsic low-$T$ value was 0.00064 eV~\cite{Li2020}.

\section{\label{Sec:MandChi} Magnetic Susceptibility and Magnetization versus Field Isotherms}

\subsection{\label{Sec:smpMHT} \smp}

\begin{figure}
\includegraphics[width=3.5in]{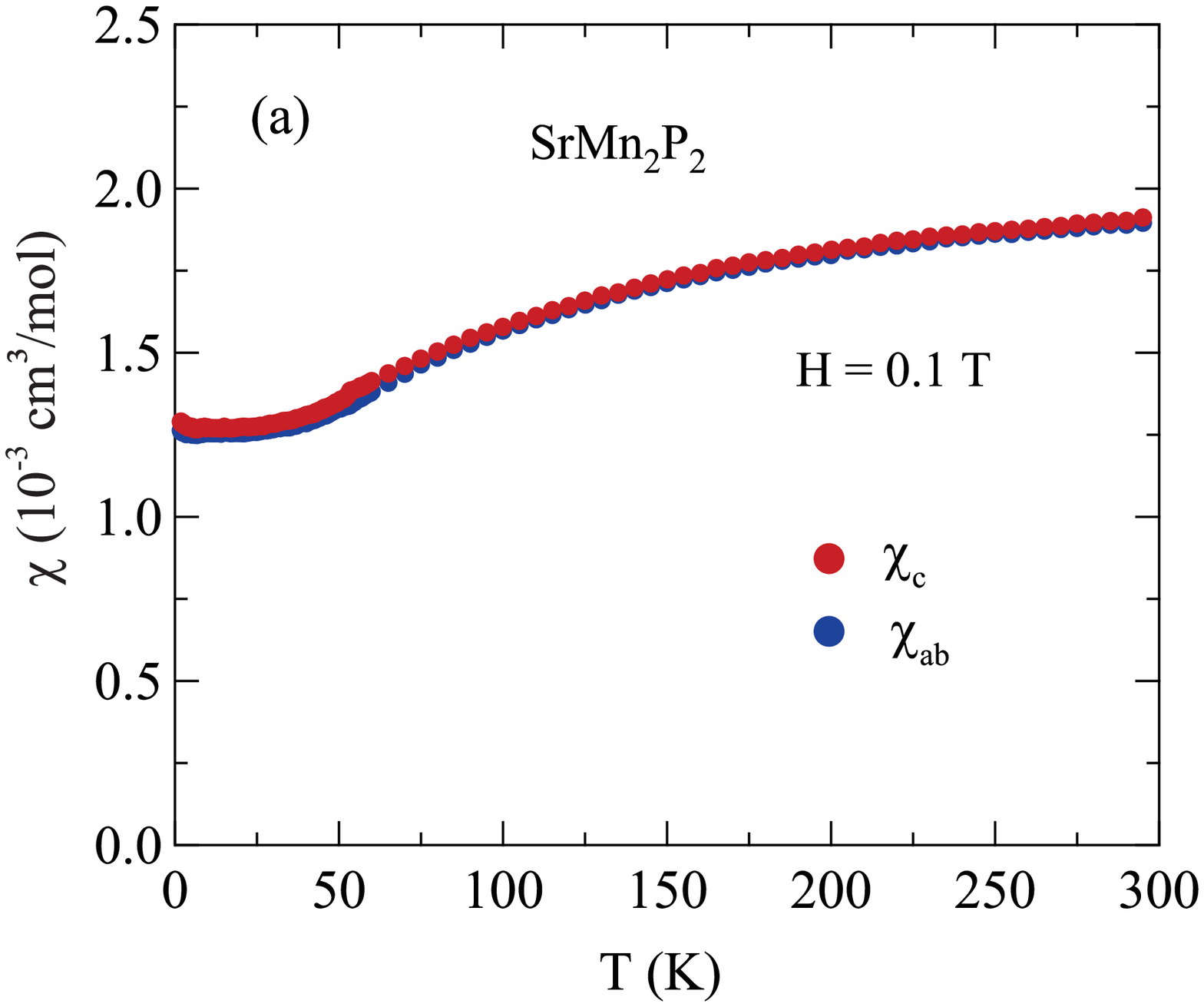}
\includegraphics[width=3.2in]{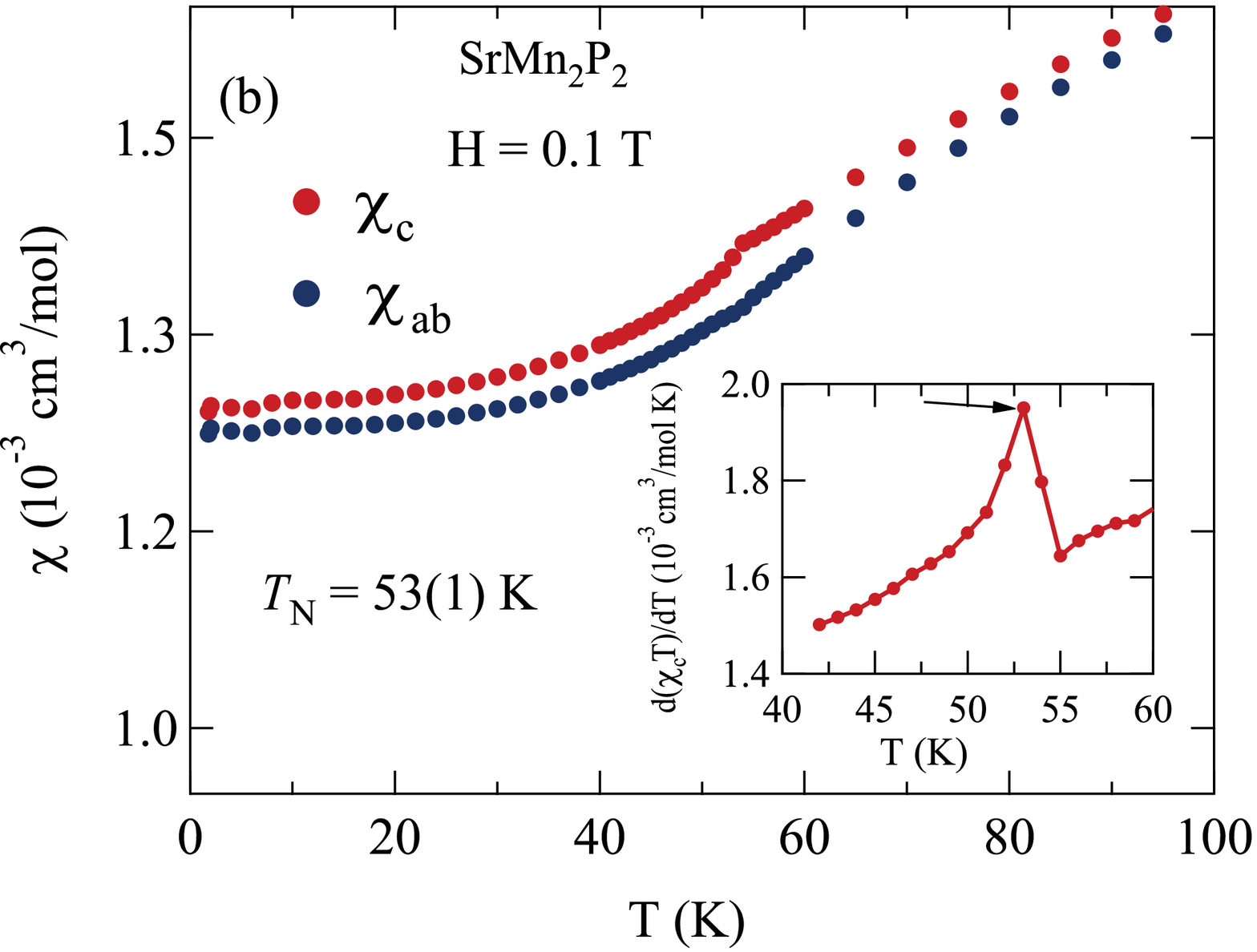}
\caption{(a)~Temperature dependent zero-field-cooled magnetic susceptibility $\chi(T)$ of \smp\ in a magnetic field $H=0.1$~T applied in the $ab$~plane ($\chi_{ab}$) and along the $c$~axis ($\chi_c$). (b) The expanded plot of $\chi(T)$ between 1.8 and 100~K to highlight the transition. Inset: Derivative $d[\chi_c(T)T]/dT$ versus $T$ for $H\parallel c$, yielding $T_{\rm N} = 53(1)$~K\@.}
\label{Fig:smpchi}
\end{figure}

\begin{figure}
\includegraphics[width=3.3in]{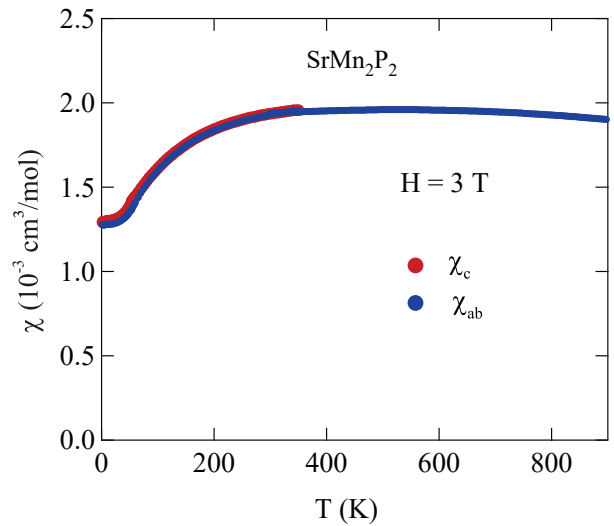}
\caption{Zero field cooled magnetic susceptibity $\chi_{ab}$ and $\chi_c$ versus $T$ for $1.8~{\rm K}\leq T \leq$ 900~K measured in $H=3$~T\@. }
\label{Fig:smpchivsm}
\end{figure}

Figure~\ref{Fig:smpchi}(a) shows the zero-field-cooled (ZFC) magnetic susceptibility $\chi(T)\equiv M(T)/H$ of \smp\ in a magnetic field $H = 0.1$~T applied in the $ab$~plane ($\chi_{ab}$) and along the $c$~axis ($\chi_c$). These data exhibit an AFM transition at $T_{\rm N} = 53(1)$~K, clearly seen from the peak in the expanded plot of $d(\chi_cT)/dT$ versus~$T$ in Fig.~\ref{Fig:smpchi}(b) according to the Fisher relation~\cite{Fisher1962}.  The $\chi(T)$ below $T_{\rm N}$ is almost isotropic and the magnetic phase transition is sharper in $\chi_c(T)$.  According to molecular field theory, the isotropic and nearly $T$-independent $\chi(T\leq T_{\rm N})$ suggests that the ordered moments form an AFM $c$-axis helix or $ab$-plane cycloidal structure with an approximately 120$^\circ$ turn angle, irrespective of the value of spin~$S$~\cite{Johnston2012, Johnston2015}. As will be shown below, $^{31}$P NMR measurements suggest an incommensurate AFM state.  Similar behaviors have been observed in $120^{\circ}$-ordered compounds such as VF$_2$ and VBr$_2$ \cite{Hirakawa1983}. The suggested magnetic structure in \smp\ is different from the magnetic structure of the isostructural compound \sma\ that was found to have a collinear AFM structure with the ordered moments aligned in the $ab$ plane, but with three approximately equally populated domains at an angle of 120$^{\circ}$ from each other~\cite{Das2017}; we found similar behavior in \sms~\cite{Sangeetha2018}. Neutron-diffraction measurements are needed confirm the magnetic structure in \smp. 

Figure~\ref{Fig:smpchivsm} shows $\chi(T)$ of \smp\ from 1.8~K to 900~K measured in $H=3$~T\@.  Unlike most local-moment antiferromagnets for which $\chi(T)$ decreases above $T_{\rm N}$ according to the Curie-Weiss law,  the $\chi(T)$ above $T_{\rm N}$ in \smp\ increases with $T$ above $T_{\rm N}$, exhibits a broad maximum at about 300~K and then slowly decreases. Therefore, the present data  suggest that strong dynamic AFM fluctuations occur up to at least 900~K, similar to corresponding data for the isostructural compounds (Sr,Ca)Mn$_2$(As,Sb)$_2$ \cite{Sangeetha2016, Sangeetha2018}. Within a local-moment picture, these features at $T>T_{\rm N}$ are characteristic of a quasi-one or -two-dimensional antiferromagnet. Similar results were obtained for a polycrystalline sample of \smp~\cite{Brock1994, Brock1994b}.

\begin{figure}
\includegraphics[width=3.3in]{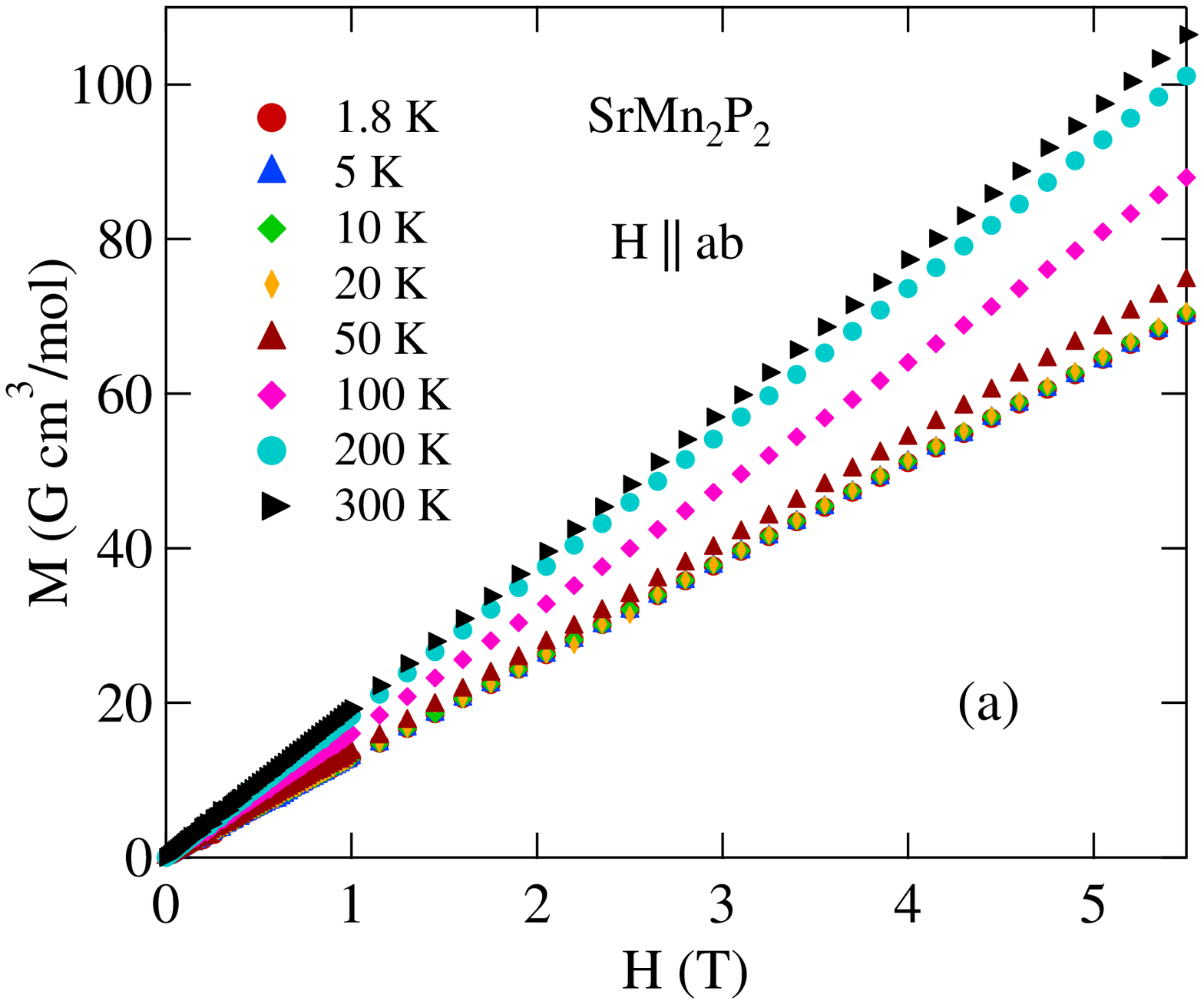}
\includegraphics[width=3.3in]{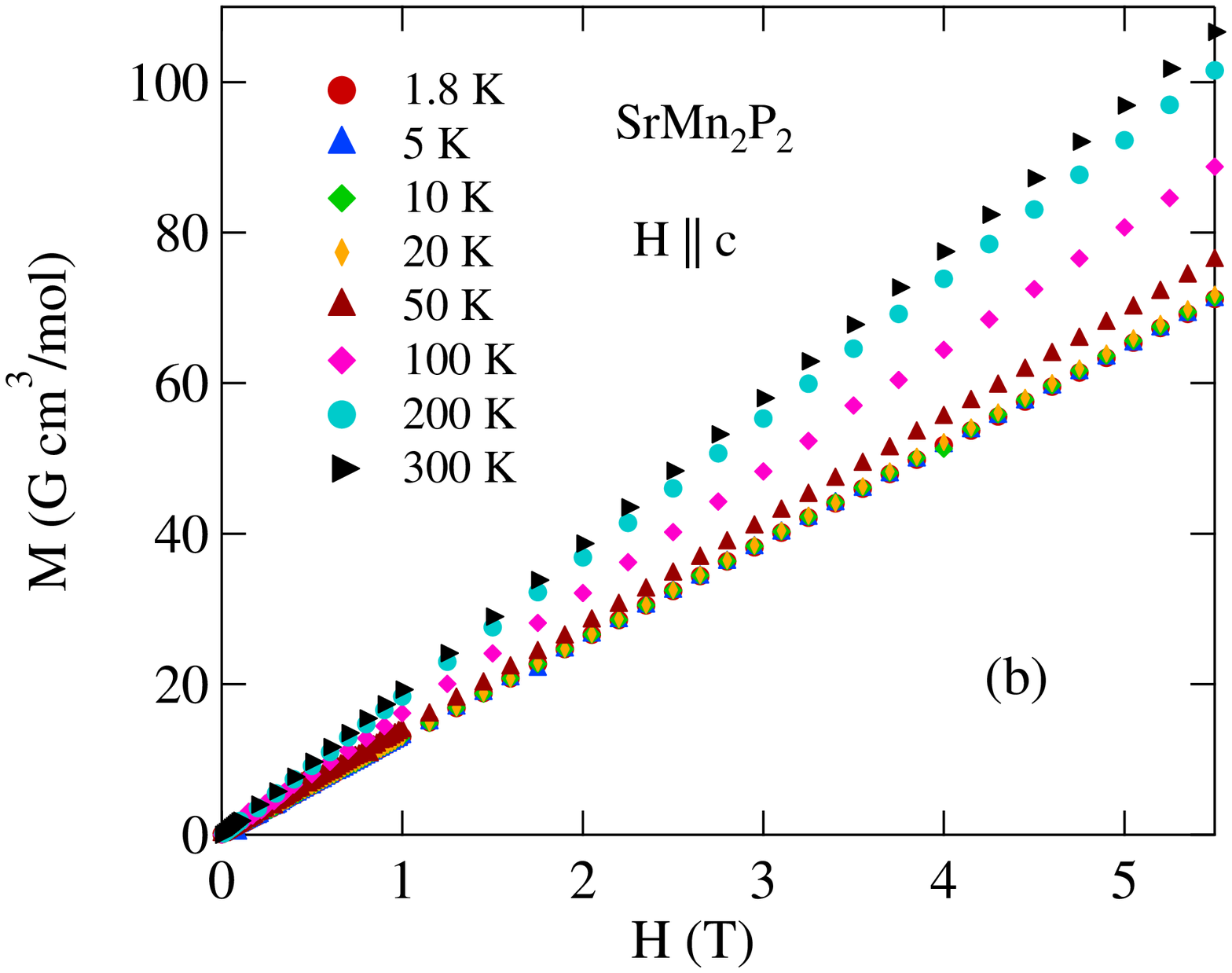}
 \protect\caption{Magnetization $M$ of \smp\ as a function of magnetic field $H$ at various temperatures $T$ with (a) $H$ in the $ab$~plane ($H\parallel  ab$) and (b) $H$ along $c$ axis ($H\parallel c$).}
\label{Fig:smpmh}
\end{figure}

$M(H)$ isotherms for a single crystal of \smp\ with $H\parallel ab$ and $H\parallel c$ are shown in Figs.~\ref{Fig:smpmh}(a) and~\ref{Fig:smpmh}(b), respectively.  The data for $M_{ab}$ and $M_c$ are proportional to~$H$ at all temperatures, indicating the absence of significant ferromagnetic or saturable paramagnetic impurities.  The nearly isotropic $M(H, T<T_{\rm N})$ data for both field directions are consistent with the nearly isotropic behavior of $\chi(T)$ in Fig.~\ref{Fig:smpchi}(a). 

We obtain an estimate of the exchange interactions between the Mn spins-5/2 in \smp\ as follows.  We assume that all spins are identical and crystallographically equivalent as in \cmp\ and \smp.  Within a local-moment Heisenberg picture, molecular-field theory predicts a magnetic susceptibility 
\bse
\bea
\chi = \frac{C}{T-\theta_{\rm p}},
\label{chiCW}
\eea
where $C$ is the Curie constant per mole of spins given by
\bea
C = \frac{N_{\rm A}g^2S(S+1)}{3k_{\rm B}},
\eea
$N_{\rm A}$ is Avogadro's number, $g$ is the spectroscopic splitting factor of a spin, and $k_{\rm B}$ is Boltzmann's constant. 
The paramagnetic Weiss temperature $\theta_{\rm p}$ is given by~\cite{Johnston2015}
\bea
\theta_{\rm p} = -\frac{S(S+1)}{3k_{\rm B}}\sum_j J_{ij},
\label{Eq:thetap}
\eea
\ese
which contains the sum of the Heisenberg exchange interactions $J_{ij}$ between a central spin~$i$ and its neighbors~$j$ with which it interacts and an AFM $J$ is positive.  Here, because a formula unit of \smp\ contains two magnetic Mn atoms, and if $C$ is expressed in cgs units of cm$^3$-K per mole of formula units, we get
\bea
C_{\rm f.u.} = \frac{2N_{\rm A}g^2S(S+1)}{3k_{\rm B}}.
\eea
Then using $g=2$ and $S=5/2$ gives
\bea
C_{\rm f.u.} = 8.75\,{\rm \frac{cm^3\,K}{mol\,f.u.}}.
\eea  

\begin{figure}
\includegraphics[width=3.4in]{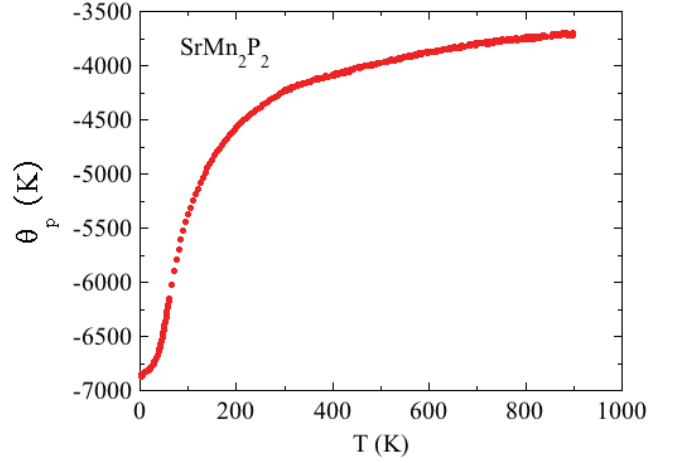}
\caption{Weiss temperature $\theta_{\rm p}$ versus temperature~$T$ obtained using Eq.~(\ref{Eq:theta(T)}) and the $\chi_{ab}(T)$ data in Fig.~\ref{Fig:smpchivsm}. }
\label{Fig:SrMn2P2_Theta_vs_T}
\end{figure}

It is clear From Fig.~\ref{Fig:smpchivsm} that $\chi_{ab}(T)$ does not attain Curie-Weiss behavior up to 900~K\@.  However,  if we calculate a $T$-dependent $\theta_{\rm p}$ from Eq.~(\ref{chiCW}) given by
\bea
\theta_{\rm p}(T) = T - \frac{C_{\rm f.u.}}{\chi(T)},
\label{Eq:theta(T)}
\eea
where $\chi(T)=\chi_{ab}(T)$ is the magnetic susceptibility per mole of formula units as in Fig.~\ref{Fig:smpchivsm}, we expect $\theta_{\rm p}(T)$ to asymptote to its actual value~$\theta_{\rm p}$ as the dynamic short-range correlations diminish with increasing $T$\@.  

Figure~\ref{Fig:SrMn2P2_Theta_vs_T} shows a plot of $\theta_{\rm p}$ versus $T$ obtained from the data in Fig.~\ref{Fig:smpchivsm} using Eq.~(\ref{Eq:theta(T)}).  The data appear to asymptote at high temperatures to a value
\bea
\theta_{\rm p} \sim -3500~{\rm K},
\label{Eq:thetapval}
\eea
where the negative sign indicates AFM interactions.  Then using Eqs.~(\ref{Eq:thetap}) and~(\ref{Eq:thetapval}) we obtain
\bea
\sum_jJ_{ij} = -\frac{3k_{\rm B}\theta_{\rm p}}{S(S+1)} \sim 0.10\,{\rm eV}.
\eea
If it is assumed that a given Mn spin only interacts with its three nearest neighbors in the corrugated honeycomb lattice with exchange interaction~$J$, then $J\sim 34\,{\rm meV}$.

\subsection{\cmp}

\begin{figure}
\includegraphics[width=3.3in]{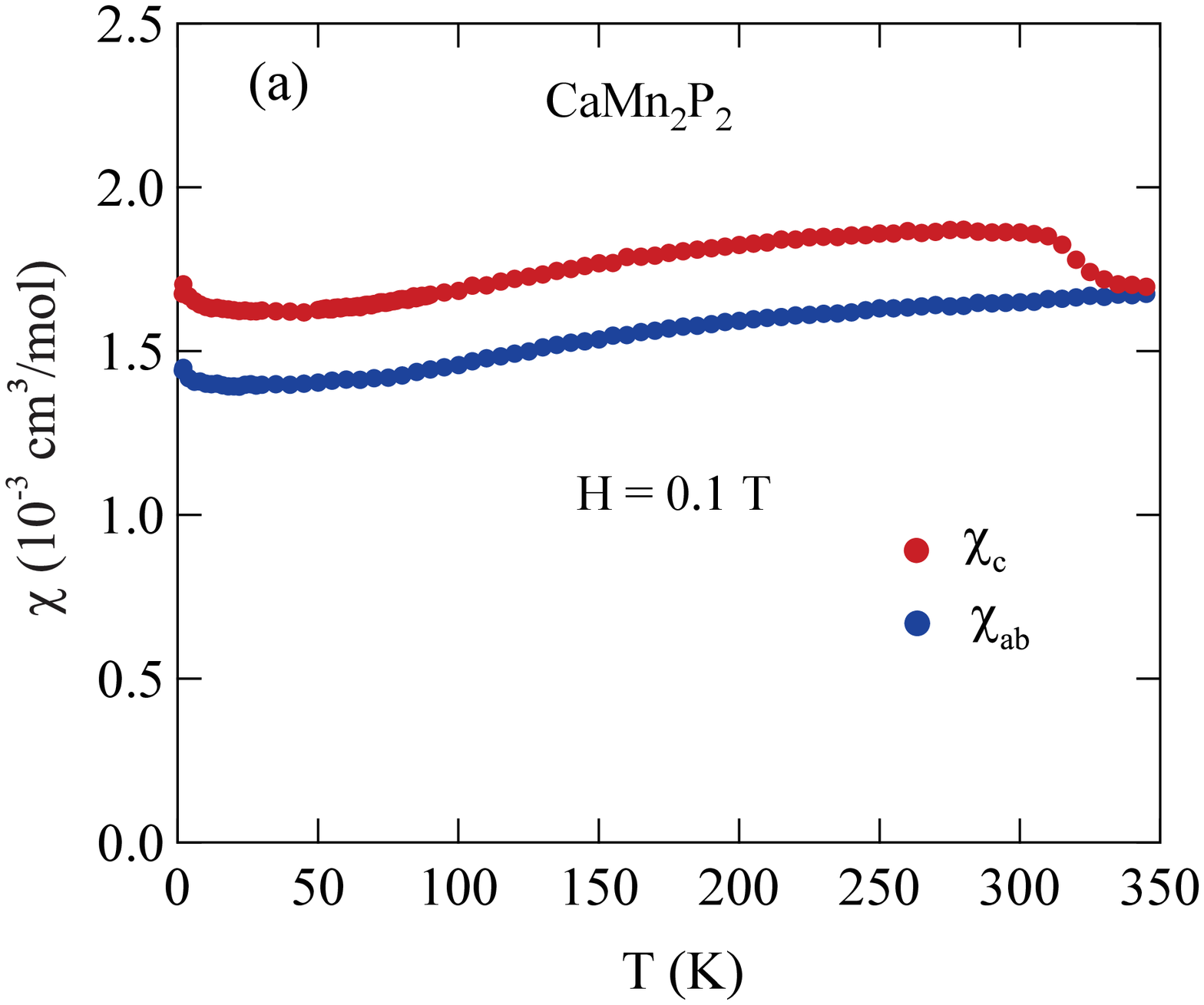}
\includegraphics[width=3.3in]{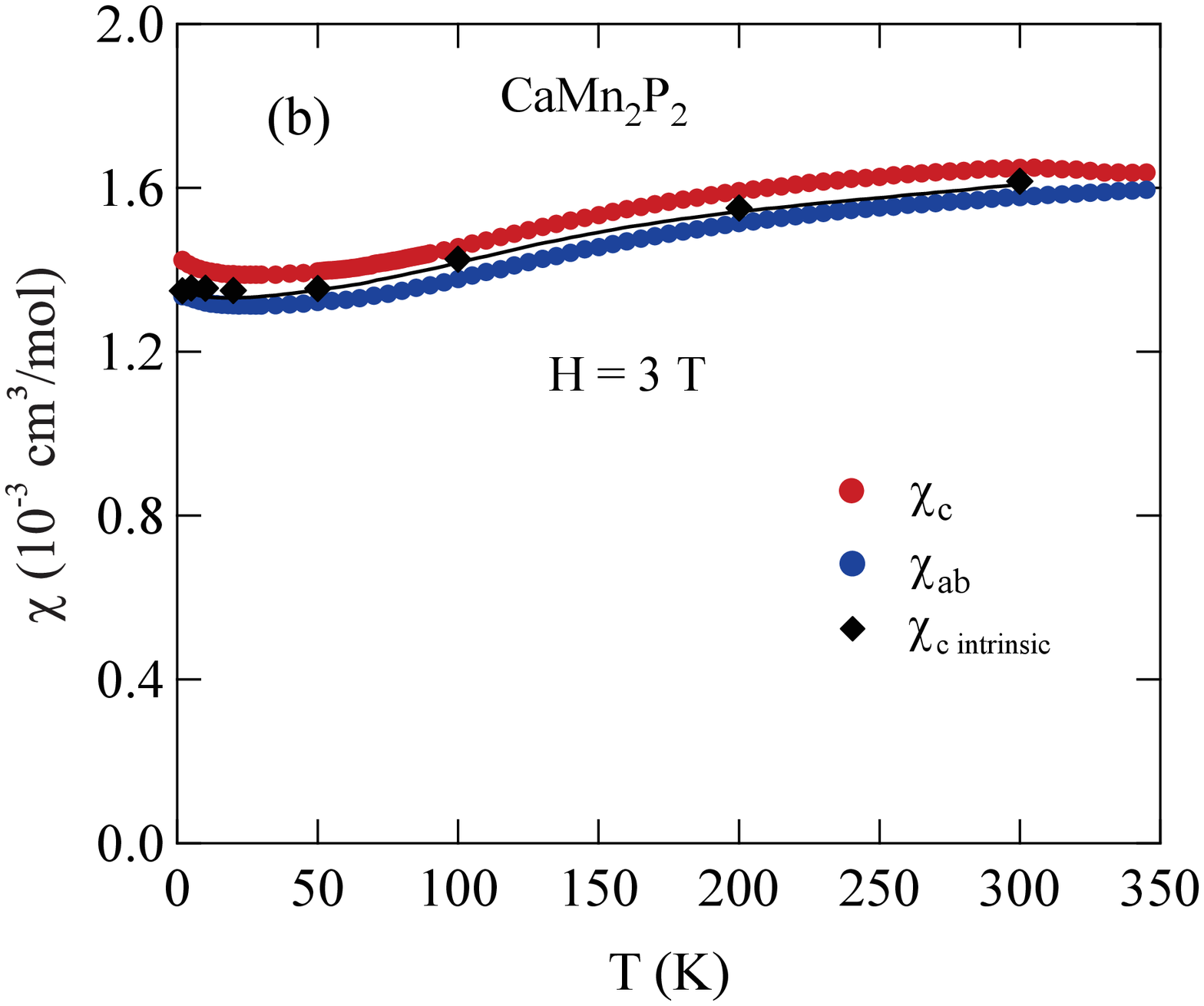}
\caption{Temperature dependent zero-field-cooled magnetic susceptibility $\chi(T)$ of \cmp\ in magnetic fields (a)~$H=0.1$~T, and (b)~$H=3$~T applied in the $ab$~plane ($\chi_{ab}$) and along the $c$~axis ($\chi_c$).  In (b), the intrinsic $\chi_c(T)$ data obtained from the high-field slope of $M(H)$ isotherms are also included, denoted as $\chi_{c\,\rm intrinsic}$.  In (a), the increase in the $\chi_c(T)$ data on cooling below 350 K is inferred to arise from a small amount of ferromagnetic MnP impurity, whereas in (b) the impurity magnetization is not as evident in the 3~T field. }
\label{Fig:cmpchi}
\end{figure}

\begin{figure}
\includegraphics[width=3.3in]{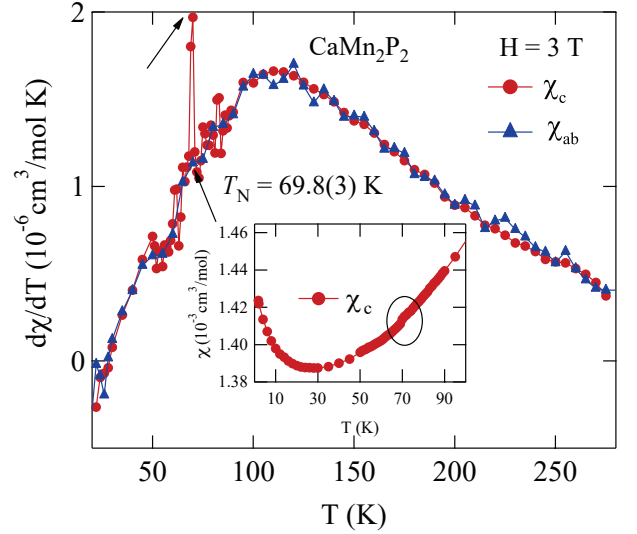}
\caption{Temperature derivative of the magnetic susceptibility $d\chi(T)/dT$ of \cmp\ in a magnetic field $H=3$~T applied in the $ab$~plane and along the $c$~axis. Inset: Expanded plot of $\chi_c(T)$ to highlight the first-order AFM transition at \mbox{$T{\rm_N}=69.8$(3)~K}\@.}
\label{Fig:cmpDchi}
\end{figure}

The zero-field-cooled (ZFC) magnetic susceptibilities $\chi \equiv M/H$ versus~$T$ measured in $H = 0.1$~T and \mbox{$H = 3$~T} applied in the $ab$~plane ($H\parallel ab,\ \chi_{ab}$) and along the $c$~axis ($H\parallel c,\ \chi_c$) for a single crystal of \cmp\ are shown in Figs.~\ref{Fig:cmpchi}(a) and~\ref{Fig:cmpchi}(b), respectively. At first glance, there is no a clear feature of any magnetic phase transition up to 350~K\@.   However, an expanded plot of the $\chi_c(T)$ data in the inset of Fig.~\ref{Fig:cmpDchi} shows a first-order AFM transition at $\approx 69$~K\@.   The temperature derivative of $\chi(T)$ versus~$T$ is plotted for both $\chi_{ab}$ and $\chi_c$ in Fig.~\ref{Fig:cmpDchi}, where the first-order nature of the magnetic transition in $\chi_c(T)$ is evident, which yields the more precise $T_{\rm N} =69.8(3)$~K\@.  Contrary to our findings, Ref.~\cite{Li2020} found that \cmp\ does not shown any long-range magnetic ordering below 400~K\@. Instead, they found a first-order structural transition at $T^*=69$~K from electrical  resistivity and heat-capacity measurements and also found that $T^*$ increases with increasing pressure.  From the nearly isotropic $\chi(T)$ below $T_{\rm N}$ in Fig.~\ref{Fig:cmpchi}(b), we suggest that the AFM structure is, or is similar to, a $120^\circ$ $c$-axis helix or $ab$-plane cycloid~\cite{Johnston2012, Johnston2015}.

Like \smp, \cmp\ is a low-dimensional antiferromagnet as seen by the very broad apparent maximum in $\chi(T)$ above $T{\rm_N}$ in Fig.~\ref{Fig:cmpchi}(b).   Indeed the Curie-Weiss temperature region of $\chi$ is not reached up to 350~K, indicating that strong AFM correlations survive to significantly higher temperatures than plotted. Previous studies of Mn pnictides suggested that for the $d^5$ electronic configuration of Mn$^{2+}$ ($S=5/2$) in trigonal Mn pnictides, the nearest-neighbor interactions are very strong and  these compounds tended to have AFM correlations or develop long-range AFM order due to the competition among different exchange interactions between the Mn sites \cite{Singh2009, Zeng2017, Sangeetha2018, Sangeetha2016}.

 In addition, a small upturn in $\chi_c(T)$ is seen in Fig.~\ref{Fig:cmpchi}(a) below about 300~K\@. This is likely due to FM MnP impurities with Curie temperature $T_{\rm c}=291.5$~K \cite{Obara1980} that are present on the crystal surface and/or as an inclusion in the crystal, similar to \bma~\cite{Singh2009} and \sma~\cite{Sangeetha2016} crystals with MnAs impurities.  From a comparison of Figs.~\ref{Fig:cmpchi}(a) and~\ref{Fig:cmpchi}(b), this FM MnP impurity is most clearly seen in the $\chi_c(T)$ data with $H=0.1$~T\@. In addition there is a small upturn in $\chi(T)$ data below $\approx 40$~K which is likely due to the contribution of paramagnetic impurities. 

\begin{figure}
\includegraphics[width=3.3in]{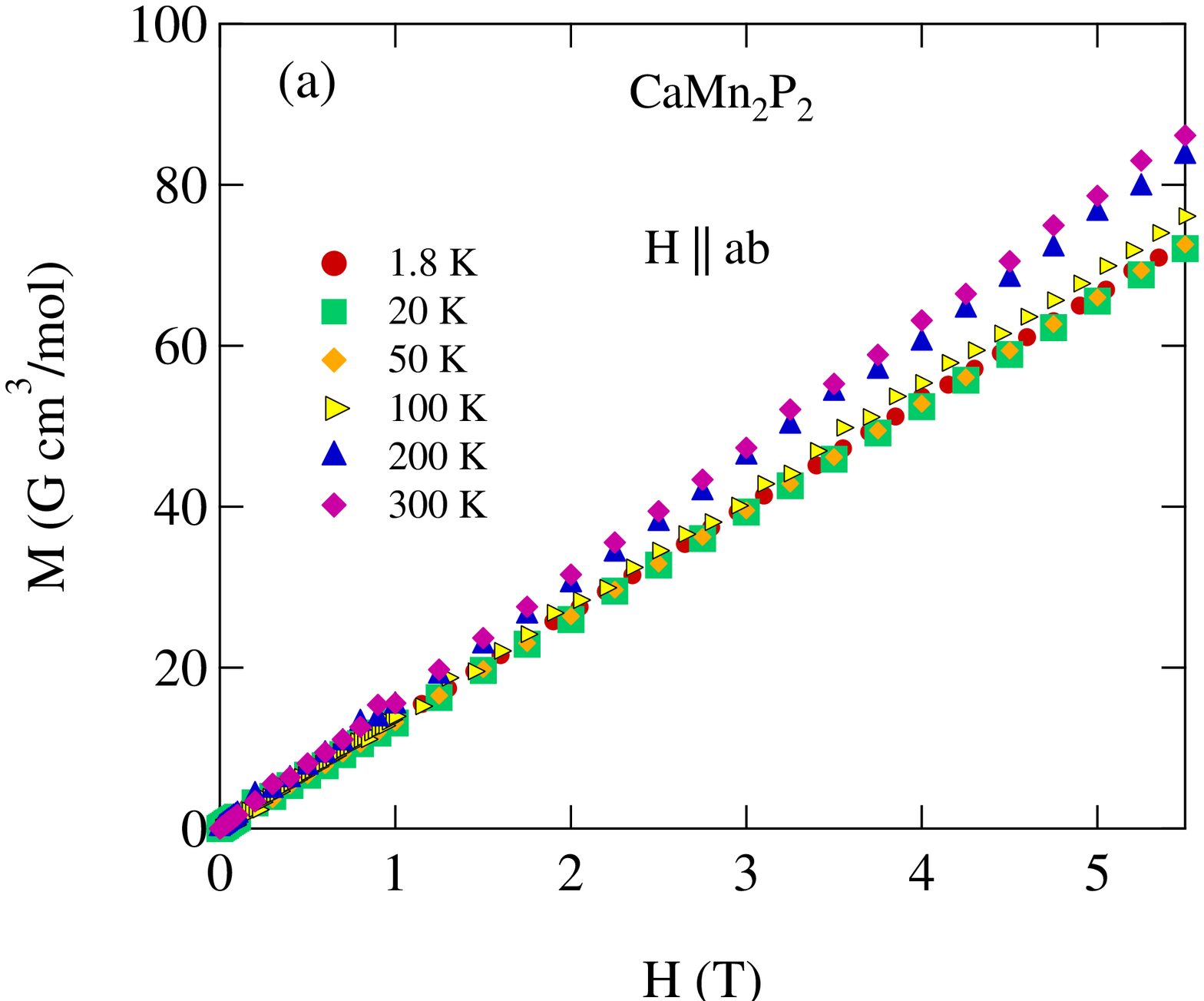}
\includegraphics[width=3.3in]{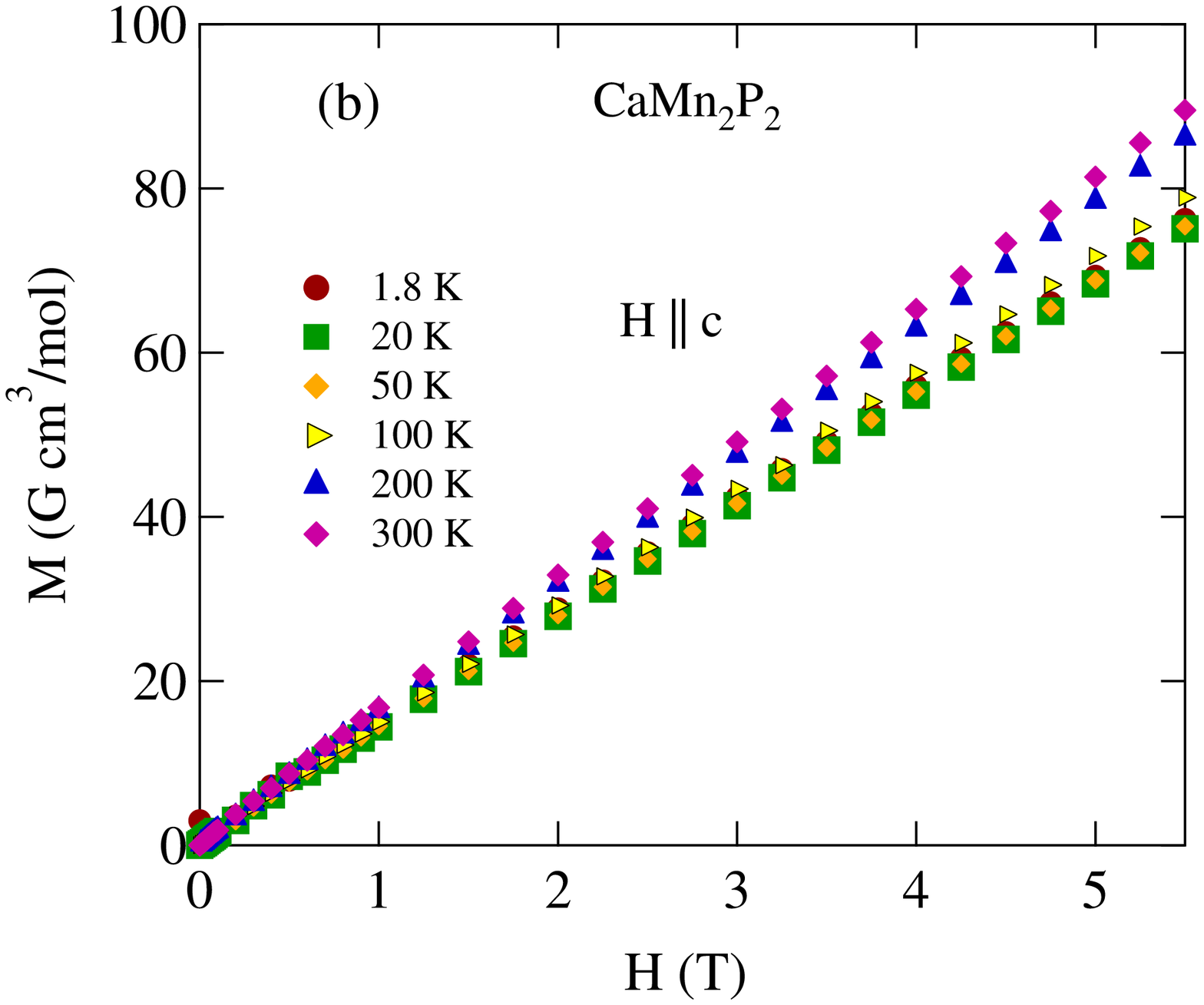}
 \protect\caption{Magnetization $M$ of \cmp\ as a function of magnetic field $H$ at various temperatures $T$ with (a) $H$ in the $ab$~plane ($H\parallel ab$~plane) and (b) $H$ along the $c$~axis ($H\parallel c$).}
\label{Fig:cmpMH}
\end{figure}

Figures~\ref{Fig:cmpMH}(a) and \ref{Fig:cmpMH}(b) show $M(H)$ isotherms for \cmp\ with $H$ in the $ab$~plane and along the $c$~axis, respectively. The $M(H)$ data are almost linear for each temperature indicating that the amount of ferromagnetic seen in Fig.~\ref{Fig:cmpchi}(a) is very small. The intrinsic magnetic behavior of $\chi_c$ data were extracted and for which the $M_c(H)$ data in the high-field range $H=3.5$--5.5~T is fitted by the linear relation
\begin{equation}
M(H,T) = M_{\rm sat}(T) + \chi_{\rm int}(T) H,
\label{eq:1}
\end{equation}
where $M_{\rm sat}(T)$ is the saturation magnetization due to the FM impurities.  The $T\to0$ value of $M_{\rm sat}$ for \cmp\ is ${\rm 2~G~cm^3/mol} = 0.00036~\mu_{\rm B}$/f.u., which corresponds to 0.03~mol\% of MnP impurities using the saturation moment $\approx 1.33~\mu_{\rm B}$/f.u.~\cite{Obara1980} for MnP\@.  The $\chi(T)\equiv M(T)/H$ data in Fig.~\ref{Fig:cmpchi}(b)  were measured with $H=3$~T\@. Therefore,  we obtained the intrinsic $\chi$ from the isotherm data according to
\begin{equation}
\chi_{\rm intrinsic}(T)=\frac{M{\rm_{measured}}(T)-M_{\rm sat}(T)}{3~\rm{T}}.
\label{Eq:Chicorrected}
\end{equation}
The $\chi_{\rm intrinsic}(T)$ data are shown by the filled black squares in Fig.~\ref{Fig:cmpchi}(b) and the black solid line is a guide to eye. It is seen that the $\chi_{\rm intrinsic}(T)$ data for \cmp\  follows the behavior of $\chi_{ab}$, which means that there is no intrinsic anisotropy between the $ab$-plane and $c$-axis magnetization data,  similar to \smp\ as seen in the previous section. 

\section{\label{Sec:HC}  Heat Capacity}

\begin{figure}
\includegraphics[width=3.3in]{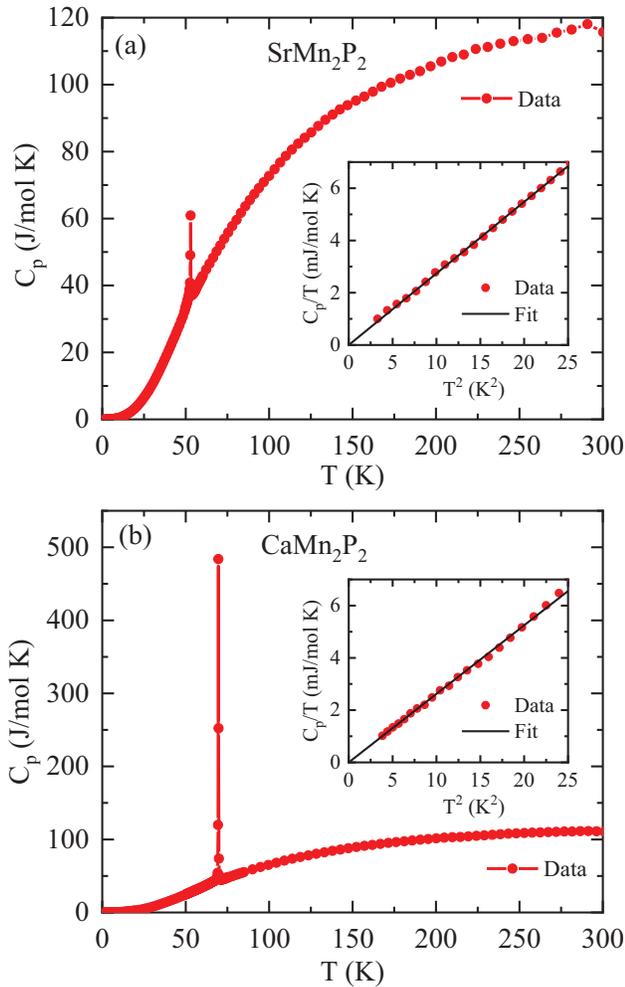}
\protect\caption{Heat capacity $C{\rm_p}$ versus temperature $T$ for (a)~\smp\ and (b)~\cmp\ single crystals. Data near $T_{\rm N}$ were obtained using the single-pulse slope-analysis method.  Insets: $C{\rm_p}(T)/T$ versus $T^{2}$ for $T\leq 5$~K, where the straight lines through the respective data are fits by Eq.~(\ref{Eq:ConT}).}
\label{Fig_heat_capacity_1}
\end{figure}

\begin{figure*}
\includegraphics[width=6.9in]{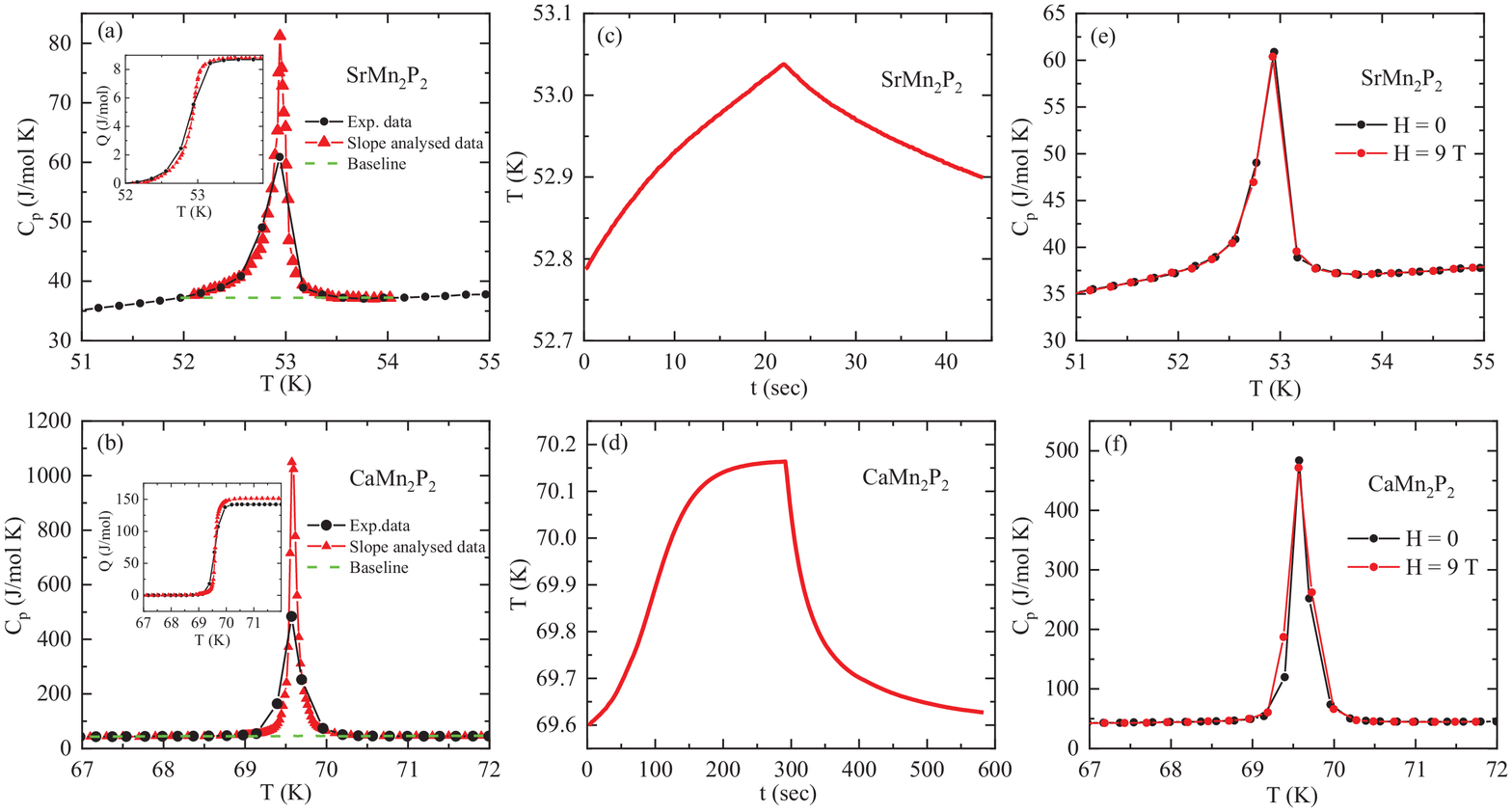}
 \protect\caption{Comparison of as-measured $C{\rm_p}(T)$ data and the single-pulse slope-analysed data (using a heating curve) for (a)~\smp\ and (b)~\cmp\ crystals. Insets: the latent heat associated with the first-order magnetic transition (FOMT). The $C{\rm_p}(T)$ values at the FOMT are  estimated by subtracting the polynomial-fitted baseline data (dashed green line) from the total $C{\rm_p}(T)$ data in that temperature region as shown in (a) and~(b). The temperature-response curve $T(t)$ at the FOMT for (c)~\smp\ and (d)~\cmp. The conventionally-measured $C{\rm_p}(T)$ is plotted for (e)~\smp\ and (f)~\cmp\ with $c$-axis applied magnetic fields $H = 0$ and 9~T, respectively.}
\label{Fig_heat_capacity_2}
\end{figure*}

Figures~\ref{Fig_heat_capacity_1}(a) and \ref{Fig_heat_capacity_1}(b) show zero-field $C{\rm_p}(T)$ data for \smp\ and \cmp, respectively. The sharp peaks in $C_{\rm p}(T)$  at 53~K in \smp\ and at 69.8~K in \cmp\ are at the respective N\'eel temperatures of the two compounds found from the above $\chi(T)$ data.  The $C{\rm_p}(T)$ value obtained at 300~K is smaller than the classical Dulong-Petit limit of $3nR \sim 124$ J/mol-K for both compounds and similar to that reported in Ref.~\cite{Li2020} for \cmp. Figures~\ref{Fig:smpchivsm} and~\ref{Fig:cmpchi} demonstrate that strong dynamic short-range AFM correlations persist up to high temperatures, as was previously found for the isostructural compounds \cma\ and \sma~\cite{Sangeetha2016}.  Thus extraction of the $T$-dependent lattice contribution to the heat capacity below 300~K which could then reveal the $T$ dependence of the magnetic contribution is not possible with the information available.

The insets of Figs.~\ref{Fig_heat_capacity_1}(a) and~\ref{Fig_heat_capacity_1}(b) show $C{\rm_p}(T)/T$ versus $T^{2}$ between 1.8 and 5~K\@. The data were fitted by
\be
\frac{C{\rm_p}}{T}=\beta T^{2}
\label{Eq:ConT}
\ee
appropriate to insulators, where $\beta$ reflects the low-$T$ lattice contribution which we assume does not contain a three-dimensional AFM spin-wave contribution. From the fits of Eq.~(\ref{Eq:ConT}) to the data in the insets of Figs.~\ref{Fig_heat_capacity_1}(a) and \ref{Fig_heat_capacity_1}(b), we obtain $\beta = 0.273(1)$~mJ/(mol~K$^{4})$ for \smp\ and 0.262(1)~mJ/(mol~K$^{4})$ for \cmp. The Debye temperature $\theta_{\rm D}$ is given by
\be
\theta{\rm_D}=\left(\frac{12\pi^{4}Rn}{5\beta}\right)^{1/3},
\label{Eq_thetaD}
\ee
where $R$ is the molar gas constant and $n$ is the number of atoms per formula unit [$n=5$ for (Sr or Ca)Mn$_2$P$_2$]. Using the above $\beta$ values we obtain $\theta{\rm_D} = 329(3)$~K and 314(1)~K for \smp\ and \cmp, respectively.

As seen in Figs.~\ref{Fig_heat_capacity_1}(a) and~\ref{Fig_heat_capacity_1}(b), in both compounds the experimentally-observed heat-capacity peak at $T_{\rm N}$ is very sharp, indicating weak and strong first-order transitions in \smp\ and \cmp, respectively.   Near $T_{\rm N}$, the respective data were obtained utilizing the single-pulse slope-analysis method with heating curves as discussed in Section~\ref{ExpDetails}. Figures~\ref{Fig_heat_capacity_2}(a) and \ref{Fig_heat_capacity_2}(b) compare the conventionally-measured and slope-analyzed $C{\rm_p}(T)$ data close to the first-order magnetic transition (FOMT) region of \smp\ and \cmp, respectively. The slope-analyzed data show higher peak values than the conventionally-measured data.

For both compounds, the peak values of $C{\rm_p}(T)$ obtained from the conventional measurement technique are seen from Fig.~\ref{Fig_heat_capacity_2} to be about a factor of two smaller than those from the slope-analyzed data. From the latter data, the height of the $C_{\rm p}$ peak for \smp\ above background is about 40~J/mol\,K, whereas that for \cmp\ is about 1000~J/mol\,K, indicating a much larger latent heat in \cmp.  This difference is clearly visible in the temperature-response versus time $T(t)$ plots around the the respective FOMT for \smp\ and \cmp\ as shown in Figs.~\ref{Fig_heat_capacity_2}(c) and~\ref{Fig_heat_capacity_2}(d), respectively.  For a FOMT, the $T(t)$ curve exhibits a plateau at the transition temperature associated with the latent heat, as clearly observed for \cmp\ at $T_{\rm N}$. Such a plateau is replaced by a region of negative curvature for \smp\ indicating a much smaller latent heat associated with the FOMT for \smp.

The latent heat $Q$ associated with the FOMT for both compounds was calculated by first subtracting the respective $C_{\rm p}(T)$ backgrounds using a polynomial fit and then measuring the area under the resultant peak. For \smp\ we obtained $Q \approx 8.7$~J/mol at $T_{\rm N} \approx 53$~K, whereas for \cmp\ we obtained $Q\approx 152$~J/mol at \mbox{$T_{\rm N} = 69.5$~K.}\@ The difference in the latent heat estimated from the conventionally-measured data and the slope-analyzed data is negligible for \smp\ and is only $\sim 11$~J/mol for \cmp\ as shown in the insets of Figs.~\ref{Fig_heat_capacity_2}(a) and~\ref{Fig_heat_capacity_2}(b), respectively.  The heat-capacity peak value in \cmp\ obtained from the slope-analyzed measurements is larger than obtained previously~\cite{Li2020}.  As shown in Figs.~\ref{Fig_heat_capacity_2}(e) and~\ref{Fig_heat_capacity_2}(f), an applied field $H=9$~T has negligible influence on $C_{\rm p}(T\approx T_{\rm N})$ for both \smp\ and \cmp, respectively.

\section{\label{NMR} Nuclear Magnetic Resonance}
 
 \subsection{SrMn$_2$P$_2$}
 
 \subsubsection{$^{31}$P NMR spectrum}

\begin{figure}[tb]
\includegraphics[width=\columnwidth]{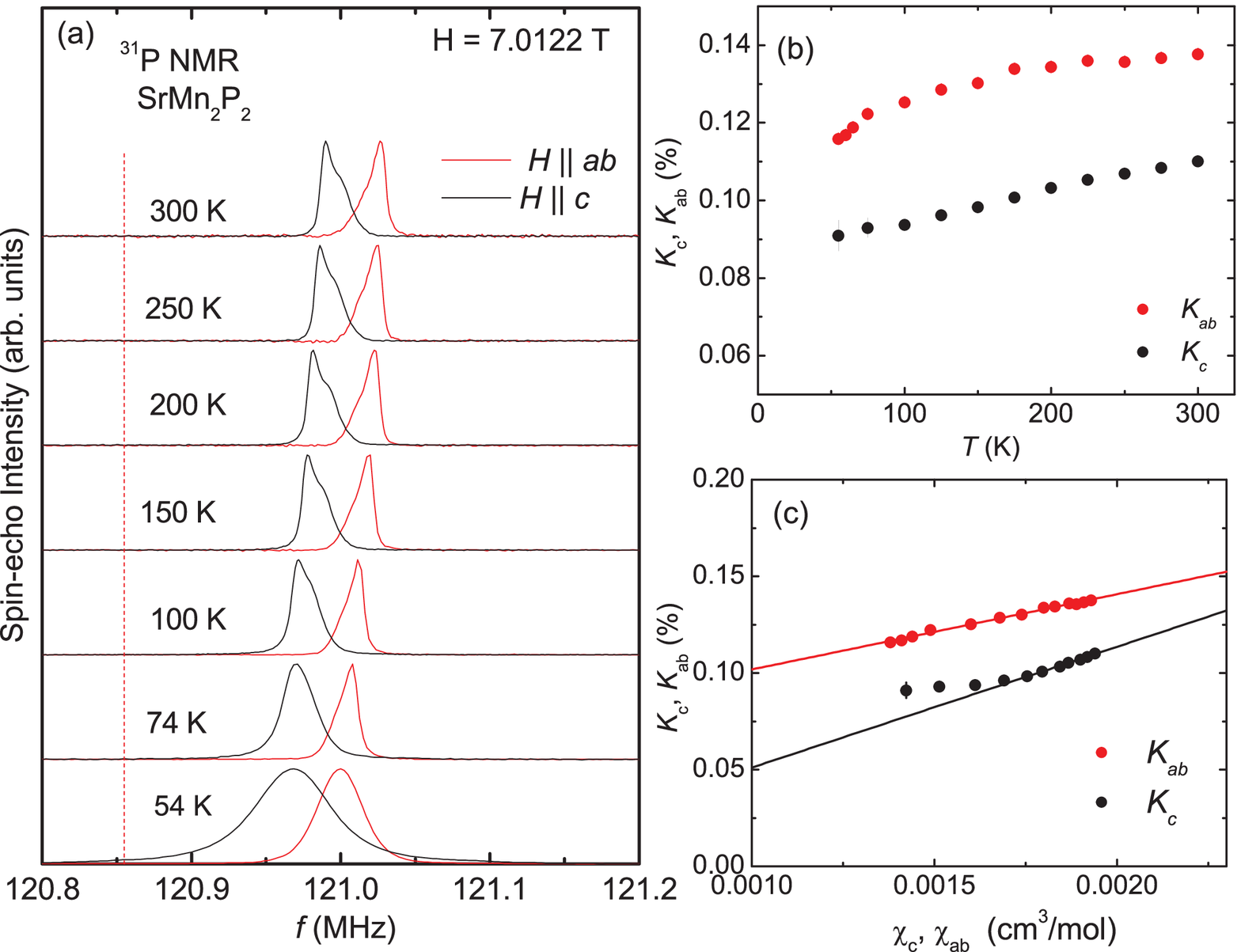} 
\caption{ (a) $^{31}$P-NMR spectra under a magnetic field \mbox{$H$ $\sim$ 7.0122 T}  parallel  to the $c$ axis (black) and parallel to the $ab$ plane (red) at various temperatures in SrMn$_2$P$_2$.  The vertical dashed line represents the zero-shift position ($K$  = 0). 
(b) Temperature dependence of  the $^{31}$P-NMR shifts $K_c$ and $K_{ab}$.
(c)  $K$ vs. $\chi$ plots for each magnetic field direction. The lines are fitting results. 
  }
\label{fig:SrMn2P2_PM}
\end{figure}   

    Figure \ref{fig:SrMn2P2_PM}(a)  shows the typical $T$ dependence of  $^{31}$P NMR spectra above $T_{\rm N}$  for $H \parallel ab$ plane (black lines) and $H \parallel  c$ axis (red lines).  
    For each magnetic field direction, we observed a single line as expected for the nuclear spin $I = 1/2$ NMR spectrum.
   However, although the NMR  line is relatively sharp at high $T$, the spectra show asymmetric shapes. 
   Since  we used a single crystal for our measurements,  the asymmetric shape indicates a slight distribution of the hyperfine field at the P sites and/or the presence of more than two P sites. 
  Since only one P site is expected from the crystal structure, the origin of the asymmetric shape of the spectra is not clear at present.
 
   Figure \ref{fig:SrMn2P2_PM}(b)  shows the $T$ dependence of the NMR shift for $H\parallel c$~axis ($K_c$) and  $H\parallel ab$~plane ($K_{ab}$) determined by the peak position for each spectrum. 
   Both $K_c$ and $K_{ab}$ decrease with decreasing $T$ as in the case of the aforementioned magnetic susceptibility. 
    The NMR shift has contributions from the $T$-dependent spin part $K_{\rm spin}$ and a $T$-independent orbital part $K_0$. 
     $K_{\rm spin}$ is proportional to the spin susceptibility $\chi_{\rm spin}$ through the hyperfine coupling constant $A$ giving $K(T)=K_0+\frac{A}{N_{\rm A}}\chi_{\rm spin}(T)$, where $N_{\rm A}$ is Avogadro's number.
     Figure \ref{fig:SrMn2P2_PM}(c)  plots $K_{ab}$ and $K_{c}$ against $\chi_{ab}$ and $\chi_{c}$, respectively, with
$T$ as an implicit parameter.  
    Here we use the magnetic susceptibilities measured at $H=3$~T\@. 
    $K_{ab}$ and $K_c$ vary with the respective $\chi$ as expected, although one can see a slight deviation from the linear relationship for $H\parallel c$. 
    The deviation could be due to the broadening of the spectrum below 100~K for $H\parallel c$.
    
    We estimated the hyperfine coupling constants \mbox{$A_{c}= (7.0 \pm 0.1)$ kOe/$\mu_{\rm B}$} and $A_{ab}= (4.4 \pm 0.1)$~kOe/$\mu_{\rm B}$ by fitting the data in Fig.~\ref{fig:SrMn2P2_PM}(c) above 100~K\@. 
    From the estimated hyperfine coupling constants, we evaluated the isotropic hyperfine coupling constant \mbox{$A_{\rm iso} = (5.23 \pm 0.10)$~kOe/$\mu_{\rm B}$} and the axially-anisotropic hyperfine coupling constant $A_{\rm ax} = (0.9 \pm 0.1)$~kOe/$\mu_{\rm B}$, respectively, from the relations $A_{\rm iso} = (A_c+2A_{ab}$)/3 and $A_{\rm ax} = (A_c-A_{ab}$)/3.
    The $A_{\rm iso}$ originates from the transferred hyperfine interactions between the $^{31}$P nucleus and its neighboring Mn$^{2+}$ ions through the mixing of the Mn~3$d$ and P~2$s$ orbitals and/or polarization of inner-core $s$~electrons via the P~2$p$ orbitals.
   On the other hand, the $A_{\rm ax}$ part comes from the dipolar interactions from the Mn$^{2+}$ spins at the P site and/or on-site dipolar field from the polarization of P 2$p$ orbitals. 
    It is noted that the hyperfine coupling constant at the P site is dominated by the isotropic part, although there is a $\sim 17\%$ contribution of the axially component with respect to the isotropic part.

\begin{figure}[tb]
\includegraphics[width=\columnwidth]{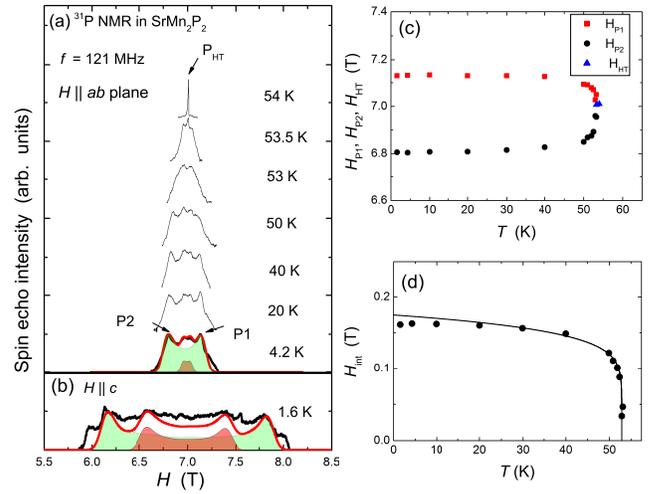} 
\caption{ (a)  Field-swept $^{31}$P-NMR spectra at a resonance frequency of $f$ = 121 MHz for $H$ parallel to the $ab$ plane at various temperatures in SrMn$_2$P$_2$ below $T_{\rm N}$. 
The black curves are the observed spectrum.
   The green and light-red areas are calculated spectra with an incommensurate helical AFM state with different internal fields of $H_{\rm int} = 2.0$ and 0.5~kOe, respectively.    
 The red curve is the sum of the two calculated spectra.
(b) Field-swept $^{31}$P-NMR spectrum at 1.6 K for $H \parallel c$.  The black curves are the observed spectra and the other colored areas and line are the same with different internal fields (see text).
(c) Temperature dependence of the peak positions for P$_{\rm HT}$, P1 and P2 defined in (a). 
(d) Temperature dependence of $H_{\rm int} = (H_{\rm P1} - H_{\rm P2})/2$. 
The solid line is the calculated result of $H_{\rm int} = H_{\rm int, 0}(1-T/T_{\rm N})^\beta$  with  \mbox{$H_{\rm int, 0} = 0.175$~T},  $\beta = 0.13$ and $T_{\rm N} = 53.0$~K\@. 
  }
\label{fig:SrMn2P2_AFM}
\end{figure}   

    When $T$ is lowered close to $T_{\rm N}\approx 53$ K, as shown in Fig.~\ref{fig:SrMn2P2_AFM}(a) for $H \parallel ab$,  a single sharp NMR line (denoted by P$_{\rm HT}$ in the figure) suddenly  broadens due to the internal field ($H_{\rm int}$) at the P site produced by the Mn$^{2+}$ ordered moments in the AFM state. The observed spectra become nearly independent of $T$ below 40 K as shown in Fig.~\ref{fig:SrMn2P2_AFM}(b), where the two peak positions denoted by P1 ($H_{\rm P1}$) and P2 ($H_{\rm P2}$) in Fig.~\ref{fig:SrMn2P2_AFM}(a) are plotted versus~$T$\@.  

    The internal field $H_{\rm int}$, which is proportional to the Mn$^{2+}$ sublattice magnetization, was determined as half of the separation between P1 and P2.
  The temperature dependences of $H_{\rm int}$ are shown in Fig. \ref{fig:SrMn2P2_AFM}(c) from which we estimated the critical exponent of the order parameter (sublattice magnetization). 
   $H_{\rm int}$ was fitted by the power law $H_{\rm int} = H_{\rm int, 0}(1-T/T_{\rm N})^\beta$  with $T_{\rm N} = 53.0$~K\@. 
   The solid line in the figure shows the curve with $H_{\rm int, 0} = 0.175$~T and   $\beta = 0.13$.
Here the value of $\beta$ is much smaller than expected for any three-dimensional magnetic material with a second-order phase transition such as $\beta = 0.33$--0.367 for 3D Heisenberg, 0.31--0.345 for 3D XY, and 0.31--0.326 for 3D Ising models \cite{Nath2009}, but close to 0.125  for the 2D Ising model \cite{Nath2009}.  These results suggest a second-order phase transition.  In addition, we were able to follow the reduction in $H_{\rm int}$ near $T_{\rm N}$ and also detected a critical-slowing-down behavior in the $T$ dependence of 1/$T_1$ around $T_{\rm N}$ (shown below).  Therefore, the magnetic phase transition for SrMn$_2$P$_2$ is considered to be characterized as a second-order phase transition.   However, the above $C_{\rm p}(T)$ data suggest a weak first-order transition, so the transition at $T_{\rm N}$ has characteristics of both orders of the transition.
On the other hand, we will show below that CaMn$_2$P$_2$ exhibits a strong first-order phase transition where we observed a clear jump in the order parameter just below the magnetic phase transition temperature and also no observation of critical slowing down in $T_1$ measurements, in contrast to our observations for  SrMn$_2$P$_2$.
 
    The broad NMR spectra observed below $T_{\rm N}$  indicate a distribution of internal fields $H_{\rm int}$ which is reminiscent of a two-horn structure expected for an incommensurate helical structure, as has been observed in EuCo$_2$P$_2$~\cite{Higa2017}  and EuCo$_2$As$_2$~\cite{Ding2017}.
   In fact,  the observed spectrum is reasonably reproduced by that calculated for an incommensurate helical AFM state shown by the green area assuming an internal field along the $ab$~plane $H_{\rm int,ab} = 2.0$~kOe, although we note that one needs to introduce another P site with a smaller $H_{\rm int, ab}$ = 0.5 kOe (shown by the light red area). 
   The red curve is the sum of the two calculated spectra.
    The origin of the two P sites is not clear at present; however, it seems to be consistent with the observed $^{31}$P NMR asymmetric spectra in the PM state which suggests that there is more than one P site having a different hyperfine coupling constant in the system. 
    It is also noted that we consider only the isotropic hyperfine field for the calculation for simplicity; therefore, the slight deviation between the calculated and observed spectra could be due to the axially anisotropic part of the hyperfine field. 

   A similar, but much broader, two-horn-like spectrum was observed for $H \parallel c$ in the AFM state as shown in Fig.~\ref{fig:SrMn2P2_AFM}(b).  
   Here we were able to measure the spectrum only at 1.6 K and not at higher~$T$ due to poor signal intensity.
   The observed spectrum was also reasonably reproduced by a calculated spectrum for the incommensurate AFM state where again we assumed two different P sites with different internal fields along the $c$ axis [$H_{\rm int,c} = 8.5$~kOe (green area), and $H_{\rm int,c} = 4.5$~kOe (light red area)].  
   As for $H\parallel ab$, the observed spectrum was not perfectly reproduced by the calculated spectrum.
  However, we consider that the analysis captures the essential point, evidencing the incommensurate AFM state in SrMn$_2$P$_2$. 
  
Our NMR results are consistent with previous $\chi(T)$ data for a polycrystalline \smp\ sample and associated powder neutron-diffraction data~\cite{Brock1994b} that suggested the presence of a complex low-dimensional~\cite{Navarro1976} incommensurate AFM structure of high-spin Mn$^{2+}$ below $T_{\rm N} = 52(2)$~K with significant short-range AFM order well above $T_{\rm N}$.
   
 \subsubsection{$^{31}$P spin-lattice relaxation rate 1/T$_1$}

\begin{figure}[tb]
\includegraphics[width=\columnwidth]{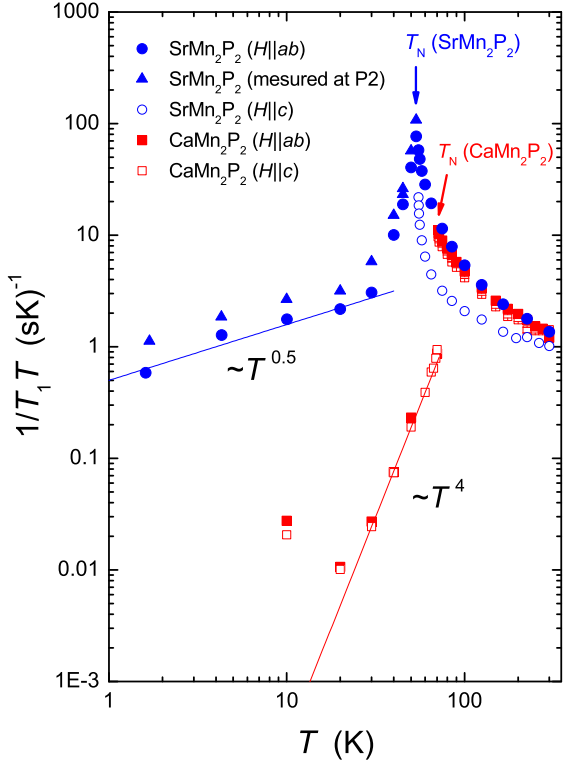} 
\caption{ Temperature dependence of 1/$T_1T$ for SrMn$_2$P$_2$ and CaMn$_2$P$_2$ for both magnetic field directions, $H \parallel c$ axis and $H\parallel ab$ plane.  
The blue circles and triangles in the AFM state of SrMn$_2$P$_2$ for $H\parallel ab$ are the results measured at nearly zero-shift and the P2 positions, respectively.  
  For the AFM state of CaMn$_2$P$_2$, 1/$T_1T$ was measured at the P3 position for $H\parallel c$ and at the lowest field peak for $H\parallel ab$.    
The blue and red straight lines show the power law dependences 1/$T_1T\propto T^{0.5}$ and $T^{4}$, respectively.
  }
\label{fig:T1}
\end{figure}   

\begin{figure}[tb]
\includegraphics[width=3.4in]{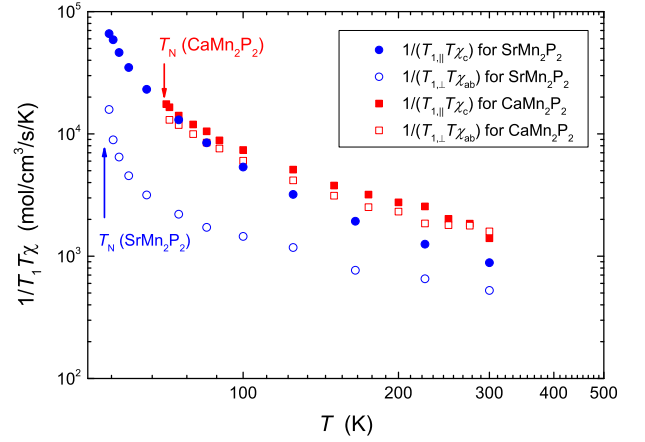} 
\caption{ Temperature dependence of $1/(T_{1,\bot}T\chi_{ab}$)  and $1/(T_{1,\|}T\chi_c$) for SrMn$_2$P$_2$ and CaMn$_2$P$_2$.  
  }
\label{fig:T1Tchi}
\end{figure}   

        Figure \ \ref{fig:T1} shows the temperature dependence of the $^{31}$P  spin-lattice relaxation rate divided by temperature 1/$T_1T$ for $H$ parallel to the $c$ axis and the $ab$ plane.
        For $H \parallel  ab$, 1/($T_1T)_{ab}$ was measured at the peak position for the spectra in the PM  state and at nearly zero shift position of the spectra in the AFM state, while  1/($T_1T)_c$ for  $H \parallel  c$  was measured only in the PM state due to poor signal intensity in the AFM state.
    
    As seen in the figure, with decreasing temperature, both 1/$T_1T$ gradually increase and strongly enhance below $\sim$ 60 K. 
  For $H\parallel ab$, 1/$T_1T$ starts to decrease just below $T_{\rm N}$ = 53 K, exhibiting a clear peak in 1/$T_1T$ which is usually explained by a critical slowing down of spin fluctuations expected for a second-order phase transition.
    Thus our 1/$T_1T$ data  show the nature of the second-order phase transition for the AFM-PM transition in SrMn$_2$P$_2$. 
 
     In order to discuss magnetic fluctuation effects in the PM state, it is useful to re-plot the $1/T_1$ data by changing the vertical axis from $1/T_1T$ to $1/T_1T\chi$ as shown in Fig.~\ref{fig:T1Tchi}. 
    In general, $1/T_{\rm 1}T$ can be expressed in terms of the imaginary part of the dynamic susceptibility $\chi^{\prime\prime}(\vec{q}, \omega_0)$ per mole of electronic spins as \cite{Moriya1963}
\bea
\frac{1}{T_1T}=\frac{2\gamma^{2}_{N}k_{\rm B}}{N_{\rm A}^{2}}\sum_{\vec{q}}|A(\vec{q})|^2\frac{\chi^{\prime\prime}(\vec{q}, \omega_0)}{\omega_0},
\eea
where the sum is over the wave vectors $\vec{q}$ within the first Brillouin zone, $A(\vec{q})$ is the form factor of the hyperfine interactions and $\chi^{\prime\prime}(\vec{q}, \omega_0)$  is the imaginary part of the dynamic susceptibility at the Larmor frequency $\omega_0$.  
    On the other hand,  the uniform $\chi$ corresponds to the real component 
 $\chi^{\prime}(\vec{q}, \omega_0)$ with $q = 0$ and $\omega_0 = 0$. 
  Thus a plot of $1/T_{\rm 1}T\chi$ versus $T$ shows the $T$ dependence of  $\sum_{\vec{q}}|A(\vec{q})|^2\chi^{\prime\prime}(\vec{q}, \omega_0)$ compared to that of the uniform susceptibility $\chi^{\prime}(0, 0)$. 
     Since $1/T_1T$ probes magnetic fluctuations perpendicular to the magnetic field~\cite{Moriya1963},  we calculated $1/(T_{1,\bot}T\chi_{ab}$) using the relation  $1/(T_{1,\bot}T$) = $1/(T_1T)_{c}$, when examining the character of magnetic fluctuations in the $ab$ plane. 
   Similarly, we estimated  $1/(T_{1,\|}T\chi_c$) for magnetic fluctuations along the $c$ axis from the relation~\cite{T1}
   \bea
   1/(T_{1,\|}T) = 2/(T_{1}T)_{ab} - 1/(T_{1}T)_{c}.
   \eea
   
     As shown in Fig.~\ref{fig:T1Tchi}, both $1/(T_{1,\bot}T\chi_{ab}$)  and $1/(T_{1,\|}T\chi_c$)  increase with decreasing $T$\@. 
    The results imply that $\sum_{\vec{q}}|A(\vec{q})|^2\chi^{\prime\prime}(\vec{q}, \omega_0)$ increases  more than \mbox{$\chi^{\prime}$(0, 0)}, evidencing a growth of spin fluctuations with $q\neq 0$.
   Thus we conclude that AFM  fluctuations exist in the PM state in SrMn$_2$P$_2$.
    It is noted that the AFM fluctuations are more enhanced along the $c$ axis than within the $ab$ plane in SrMn$_2$P$_2$.  
    It is also interesting to point out that from the smooth extrapolation of the $T$ dependence of both $1/(T_{1,\bot}T\chi_{ab}$)  and $1/(T_{1,\|}T\chi_c$),  the AFM fluctuations seem to persist up to temperatures much higher than 300 K, consistent with the $\chi(T)$ results discussed above.

     In the AFM state below $T_{\rm N}$, 1/$T_1T$ for $H\parallel ab$ decreases slowly where 1/$T_1T$ shows a $T^{0.5}$ power-law behavior.  
    We also measured 1/$T_1T$ in the AFM state at the lower field peak (P2) which shows similar power-law behavior with slightly different values (shown by the blue triangles in Fig.~\ref{fig:T1}).
    In the AFM state,  1/$T_1T$ is mainly driven by scattering of magnons, leading to $T^2$ and $T^4$  power-law $T$ dependencies due to a two- or three-magnon Raman process, respectively~\cite{Beeman1968}. 
  The weak $T$ dependence of 1/$T_1T\propto T^{0.5}$  below 40~K cannot be explained by the magnon scattering, and suggests the presence of other magnetic fluctuations in the magnetically-ordered state.

 \subsection {CaMn$_2$P$_2$}
 \subsubsection{$^{31}$P NMR spectrum}

\begin{figure}[tb]
\includegraphics[width=\columnwidth]{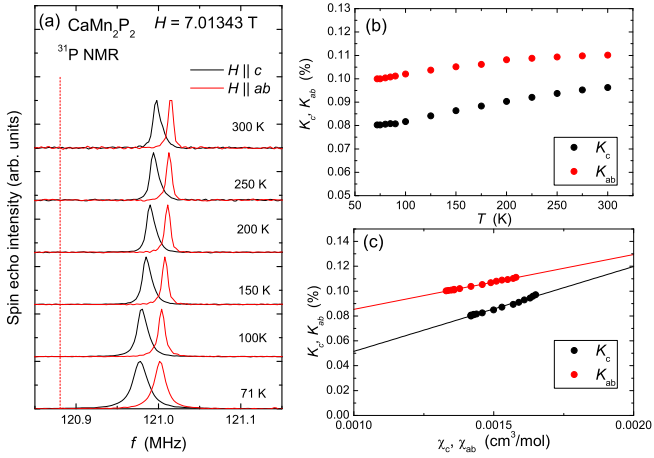} 
\caption{(a)~ $^{31}$P-NMR spectra under a magnetic field $H$ = 7.01343~T  parallel  to the $c$~axis (black) and parallel to the $ab$~plane (red) at various temperatures in CaMn$_2$P$_2$.  The vertical dashed line represents the zero-shift position ($K  = 0$). 
(b) Temperature dependence of  the $^{31}$P-NMR shifts $K_c$ and $K_{ab}$.
(c)  $K$ vs. $\chi$ plots for each field direction. The lines are fitting results.
  }
\label{fig:CaMn2P2_PM}
\end{figure}   

   Similar $^{31}$P-NMR measurements were performed on CaMn$_2$P$_2$.
   Figure~\ref{fig:CaMn2P2_PM}(a) shows the typical $T$ dependence of NMR spectra for $H \parallel  ab$ and $H \parallel  c$ in the PM state.
   As in the case of SrMn$_2$P$_2$, a single NMR line was observed. 
   The $T$ dependencies of the NMR shifts are shown in Fig.~\ref{fig:CaMn2P2_PM}(b), and Fig.~\ref{fig:CaMn2P2_PM}(c) shows  $K$-$\chi$ plots for both magnetic field directions.
   From the slopes in Fig.~\ref{fig:CaMn2P2_PM}(c), the hyperfine coupling constants were estimated to be $A_{c}= (7.6 \pm 0.1) $ kOe/$\mu_{\rm B}$ and $A_{ab}= (4.9 \pm 0.1)$\,kOe/$\mu_{\rm B}$ for $H \parallel c$ and $H \parallel ab$, respectively, leading to 
 the isotropic hyperfine coupling constant $A_{\rm iso} = (5.8 \pm 0.1)$\,kOe/$\mu_{\rm B}$ and the axially-anisotropic hyperfine coupling constant $A_{\rm ax} = (0.9 \pm 0.1)$\,kOe/$\mu_{\rm B}$.

\begin{figure}[tb]
\includegraphics[width=\columnwidth]{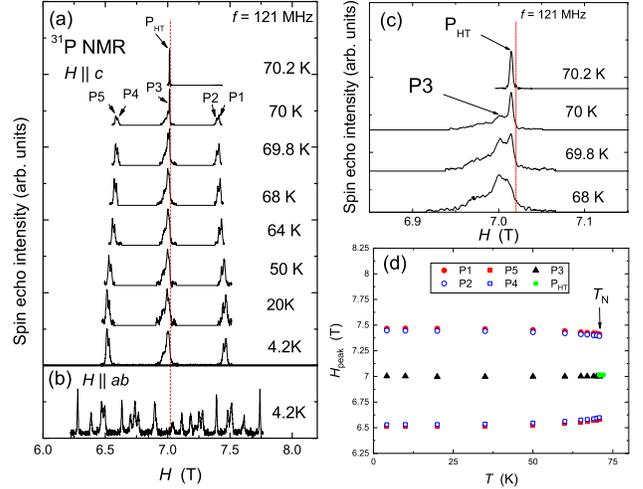} 
\caption{(a)  Field-swept $^{31}$P-NMR spectra at a resonance frequency $f = 121$~MHz for $H\parallel c$  in CaMn$_2$P$_2$ at various $T$ below $T_{\rm N}$. 
(b) Field-swept $^{31}$P-NMR spectrum at 4.2 K for $H \parallel  ab$. 
(c) Expanded $^{31}$P NMR spectra near the zero-shift position for temperatures near $T_{\rm N}$.
(d) Temperature dependence of the peak positions for P$_{\rm HT}$ and P1--P5  defined in~(a).}
\label{fig:CaMn2P2_AFM}
\end{figure}

   Although the $^{31}$P NMR spectra in the PM state and the values of the hyperfine coupling constants in CaMn$_2$P$_2$ are similar to those observed in SrMn$_2$P$_2$, surprisingly, the $^{31}$P NMR spectra in the AFM state below $T_{\rm N}$ are quite different from the case of  SrMn$_2$P$_2$.
   As shown in Fig. \ref{fig:CaMn2P2_AFM}(a) for $H \parallel c$,  the single NMR line observed in the PM state  suddenly splits into mainly three lines at the higher and lower magnetic field positions with a double-peak structure  (denoted by P1, P2, P4, and P5) and around a nearly zero-shift position (P3) in the AFM state.
   The detailed change in the NMR spectrum around zero-shift position are shown in  Fig. \ref{fig:CaMn2P2_AFM}(c). 
   Note here we measured the spectrum with increasing $T$\@. 
   No $P_{\rm HT}$ signal from the PM state could be observed and only the signal (P3) from the AFM state was detected at 68 K. 
  Then, $P_{\rm HT}$ starts to appear at 69.8 K and the P3 signal disappears completely at 70.2~K\@.  
  The coexistence of the two signals from the AFM  and PM states  can be seen in a quite narrow temperature range from 69.8 K to 70 K, indicating a very small hysteresis. 
  
   The distinct splittings of the NMR line below $T_{\rm N}$ clearly indicate that the AFM state is commensurate, which is  in strong contrast to the case of the incommensurate AFM state in SrMn$_2$P$_2$.
    Similar distinct splittings of NMR lines  were also observed for $H \parallel  ab$ where a more complicated spectrum with at least 20 peaks were detected [see, Fig. \ref{fig:CaMn2P2_AFM}(b)].
   Although the spectrum is complicated and suggests a complex magnetic structure, it is clear that the commensurate nature holds not only along the $c$ axis but also in the $ab$ plane.
    
   Figure~\ref{fig:CaMn2P2_AFM}(d) shows the $T$ dependence of the peak positions for $H\parallel c$. 
   Clear jumps in the positions due to a finite internal field below $T_{\rm N}$ can be seen. 
   These results clearly show that the AFM phase transition in CaMn$_2$P$_2$ is of first order, consistent with the results of the $C_{\rm p}(T)$ measurements.

 \subsubsection{$^{31}$P spin-lattice relaxation rate 1/T$_1$}
 
    The clear nature of the first-order phase transition in CaMn$_2$P$_2$ can be also detected in the temperature dependence of 1/$T_1T$.
    As shown in Fig.~\ref{fig:T1}, in the PM state,  1/$T_1T$ for $H\parallel c$ and $H\parallel ab$ gently increases with decreasing $T$ with no obvious anisotropy. 
    1/$T_1T$ for both magnetic field directions in CaMn$_2$P$_2$ does not exhibit a clear enhancement close at $T_{\rm N}$, indicating no critical slowing down of the Mn spins as expected for the first-order phase transition.
   In addition, 1/$T_1T$ shows a discontinuous decrease just below $T_{\rm N}$, which again confirms the first-order nature of the AFM magnetic phase transition in CaMn$_2$P$_2$.

    The increases of 1/$T_1T$ in the PM state are due to the growth of AFM fluctuations which persist to much higher temperatures above 300~K as in the case of SrMn$_2$P$_2$.
    However, as shown in Fig. \ref{fig:T1Tchi}, $1/(T_{1,\bot}T\chi_{ab}$)  and $1/(T_{1,\|}T\chi_c$) are nearly the same which suggests nearly isotropic AFM fluctuations in the PM state of CaMn$_2$P$_2$, in contrast to the case of SrMn$_2$P$_2$.

   In the AFM state, the 1/$T_1T$ data shown in Fig. \ref{fig:T1} were measured at the P3 position for $H\parallel c$ and at the lowest-field peak for $H\parallel ab$. 
  We also measured 1/$T_1T$ at different peak positions for both magnetic field directions and found no obvious difference in the values of 1/$T_1T$\@. 
   1/$T_1T$ for $H\parallel c$ and $H\parallel ab$ show $T^{4}$ power-law behaviors which are consistent with $T^4$ expected for the three-magnon relaxation process in AFM materials where the deviation from the power-law behavior  for $T \lesssim 20$~K could be due to relaxation associated with impurities.

\section{\label{Sec:Summary} Concluding Remarks}

Single crystals of \smp\  and \cmp\ have been grown using Sn flux and characterized by single-crystal x-ray diffraction, electrical resistivity $\rho$, heat capacity $C_{\rm p}$, and NMR measurements versus temperature $T$, and anisotropic magnetic susceptibility $\chi$ and magnetization $M$ versus $T$ and applied magnetic field $H$ measurements. Room-temperature single-crystal x-ray diffraction measurements confirm that both \smp\ and \cmp\ adopt the trigonal \cas-type structure containing corrugated honeycomb quasi-two-dimensional Mn spin lattices as previously reported. The $\rho(T)$ measurements demonstrate insulating ground states for both compounds with intrinsic activation energies of 0.124 eV for \smp\ and 0.088 eV for \cmp\@. The $\chi(H,T)$ and $C\rm_p(T)$ measurements reveal first-order AFM transitions at $T\rm_N$ = 53(1)~K and 69.8(3)~K for \smp\ and \cmp, respectively. 

Li and coworkers reported a first-order transition at 69.5~K at in \cmp\ from $C_{\rm p}(T)$ and $\rho(T)$ measurements in 2020~\cite{Li2020} and from Raman scattering measurements inferred that it was related to a structural transition such as superstructure formation although its potential concomitant magnetic character was not identified.

First-order AFM transitions in $H=0$ are rather unusual.  A first-order magnetostructural transition was observed at $T_{\rm N} = 205$~K  in the body-centered-tetragonal metallic 122-type Fe-based pnictide ${\rm SrFe_2As_2}$ which exhibits a transition  to an orthorhombic structure with commensurate, collinear, itinerant, spin-density-wave order~\cite{Krellner2008, Zhao2008, Jesche2008, Tam2019}.  This is a different class of materials from \cmp\ and \smp\ which are electrical insulators.  The cubic pyrite-structure insulator MnS$_2$ containing Mn$^{2+}$ cations with high-spin $S = 5/2$ and (S$_2)^{2-}$ species was found from neutron-diffraction measurements to exhibit a first-order AFM transition at $T_{\rm N} = 47.7$~K~\cite{Hastings1976, Chattopadhyay1984}. Other examples of materials exhibiting first-order AFM transitions include insulating UO$_2$ with $T_{\rm N} = 30.8$~K~\cite{Frazer1965} and MnO~\cite{Shull1951, Bloch1974} with $T_{\rm N} \approx 120 $~K, and Cr and Eu metals.  The AFM structure in Cr metal below $T_{\rm N}=311$~K is an itinerant spin-density wave~\cite{Fawcett1988}, whereas it is a \mbox{$S=7/2$} local-moment helical AFM state below $T_{\rm N} \approx 90$~K in Eu metal~\cite{Nereson1964, Gerstein1967, Cohen1969}.

In a series of papers, it was found  that symmetry considerations and renormalization-group theory could determine whether or not a given material would exhibit a first-order transition at its N\'eel temperature~\cite{Bak1976A, Mukamel1976A, Mukamel1976B, Bak1976B}.  The theory correctly predicted the occurrence of first-order transitions in the above materials UO$_2$, MnO, Cr, and Eu at their respective N\'eel temperatures.  It would be interesting to see if the same theory would predict the observed first-order AFM transitions in \cmp\ and \smp.  If not, then a structural transition at $T_{\rm N}$ would evidently be required to explain the observed first-order transitions.  For example, a crystal-structure distortion at $T_{\rm N}$ was stated to result in a first-order AFM transition in GeNCr$_3$~\cite{Zu2016}.

\acknowledgments

This research was supported by the U.S. Department of Energy, Office of Basic Energy Sciences, Division of Materials Sciences and Engineering.  Ames Laboratory is operated for the U.S. Department of Energy by Iowa State University under Contract No.~DE-AC02-07CH11358.


\begin{thebibliography}{60}
\bibitem{T1} Strictly speaking, this simple relation can not be applicable if 1/$T_1T$ has a strong anisotropy in the $ab$ plane. For CaMn$_2$P$_2$, similar values of 1/($T_1T)_c$ and 1/($T_1T)_{ab}$ suggest no strong anisotropy in  1/$T_1T$ within the $ab$ plane so that one can use the relation. In the case of SrMn$_2$P$_2$, since 1/$T_{1,\parallel}T$ is much greater than 1/$T_{1,\bot}T$, the relation can still be applicable.
\end{thebibliography}
\end{document}